%% file: ms_revised.tex
\documentclass[12pt,preprint]{aastex}

\usepackage{graphicx}
\usepackage{amssymb}
\usepackage[figuresright]{rotating}

\slugcomment{To appear in Ap. J.}

\shorttitle{Spitzer H$_2$ emission in protostellar outflows} \shortauthors{Giannini
et al.}

\begin{document}


\title{Spitzer spectral line mapping of protostellar outflows: III - H$_2$ emission in L1448, BHR71, and NGC2071}

\author{Teresa Giannini$^1$, Brunella Nisini$^1$, David Neufeld$^2$, Yuan Yuan$^2$, Simone Antoniucci$^1$, and Antoine Gusdorf$^3$ \\
}
\altaffiltext{1}{INAF - Osservatorio Astronomico di Roma, via Frascati 
33, 00040 Monte Porzio Catone, Italy}
\altaffiltext{2}{Department of Physics and Astronomy, John Hopkins University, 3400 North Charles Street, Baltimore, MD 21218}
\altaffiltext{3}{Max Planck Institut f\"ur Radioastronomie, Auf dem H\"ugel, 69, 53121 Bonn, Deutschland}



\begin{abstract}
{\it Spitzer}-IRS maps of H$_2$ pure rotational lines from S(0) to S(7) in three outflows from Class 0 sources, namely L1448,
BHR71, and NGC2071, are presented. These lines are used, in conjunction with available rovibrational, near-infrared H$_2$ lines, to probe the physical conditions of the warm gas between hundreds and thousands of Kelvin. We have constructed maps of the molecular hydrogen column density, ortho-to-para ratio and volume density, together with the index $\beta$ of the power law describing the distribution of gas temperature.
In all three outflows, the present ortho-to-para ratio significantly deviates from the high temperature equilibrium of 3, being on average between 2.0 and 2.3. These low values, that in general reflect the young age of these flows, are found also in regions of relatively high temperature (around 1000 K), likely indicating that shocks are occurring in a time shorter than that needed for a complete
para to ortho conversion. Density maps indicate upper limits close to LTE conditions, i.e. between 10$^6$-10$^7$ cm$^{-3}$; moreover we demonstrate,
on the basis of the detections of HD emission spots (R(3) and R(4) lines) that a density stratification does exist, with the low density components
(10$^4$-10$^5$ cm$^{-3}$) associated to the coldest gas. The $\beta$ index is found in all flows to be above 3.8: this value is consistent with predictions of multiple C-type bow shocks with a range of velocities, some of which insufficient to achieve the temperature at which H$_2$ partially dissociates. The contribution of H$_2$ to the total cooling is quantitatively similar to that of other abundant molecules emitting in the far-infrared, such as water and CO; moreover, the
luminosity radiated away is comparable with the estimated kinetic energy of the swept-out outflow: this supports the scenario in which the shock from which H$_2$ is emitted is also capable of accelerating the molecular outflow driven at the shock working surface. 
\end{abstract}

 

\section{Introduction}

The physical effect of jets from accreting protostars on the parental cloud is the formation of shock waves. The accelerated, compressed and heated gas radiates away the accumulated thermal energy through the emission of lines from various species, in different proportions that depend on the relative abundances and the specific shock conditions. In this respect,  
a fundamental r\^ole is played by the molecular 
hydrogen, whose high abundance compensates for the fact that its homonuclear nature leads to quadrupole-allowed rovibrational transitions with small radiative rate. Rovibrational H$_2$ lines in the near-infrared, and in particular the 1-0\,S(1) line at 2.12\,$\mu$m, have been largely used to discover jets from young protostars, and, at present around 1000 individual regions 
of shock emission have been identified (Davis et al. 2010). However, the most efficient cooling channels of molecular hydrogen are the pure rotational lines of the ground vibrational level, lying in the mid-infrared at $\lambda \le$ 28 $\mu$m (e.g. Kaufman \& Neufeld 1996). These lines represent a key tool to penetrate the very obscured regions that surround the so-called Class 0 sources, i.e. the very young protostars in the main accretion phase of their evolution (age $\sim$ 10$^4$ yr, Andr\'e, Ward-Thompson \& Barsony 1993); moreover the H$_2$ mid-infrared lines  are easily excited at few hundreds of Kelvin, and consequently they represent a valuable tool for tracing the cold gas component in the jet, where the initial conditions of the emitting gas are often still manifested in a non equilibrium value of the H$_2$ ortho-to-para ratio (Timmermann 1998, Wilgenbus et
al. 2000, Maret et al. 2009, Barsony et al. 2010). In addition to the different excitation energies, the pure rotational lines differ from the rovibrational ones for being characterized by lower critical densities: all these circumstances make the combination of the two sets of lines as particularly valuable for deriving a global view of gas excitation in molecular outflows and for inferring the total H$_2$ cooling, in particular for the environments of the Class 0 protostars.\\
The H$_2$ pure rotational lines were first observed extensively with the {\it Infrared Space Observatory} (ISO), by means of the {\it ISOCAM} camera and the {\it SWS} spectrograph.
These studies have demonstrated the usefulness of the H$_2$ rotational lines for deriving fundamental parameters of molecular shocks, such as temperature, column density, and ortho-to-para ratio, although averaged on large areas due to the poor spatial resolution of {\it ISO} (Noriega-Crespo, Garnavich \& Molinari 1998, Neufeld, Melnick \& Harwit 1998, Smith, Eisl\"offel \& Davis 1998, Nisini et al. 2000, Molinari et al. 2000,  Rosenthal, Bertoldi \& Drapatz 2000, Benedettini et al. 2000, Lefloch et al. 2003). More recently, the {\it IRAC} camera (Fazio et al. 2004) and the {\it IRS} spectrograph (Houck et al. 2004) onboard {\it Spitzer} (Werner et al. 2004), have allowed us to expand upon the handful of outflows observed with  {\it ISO} (e.g. Smith et al. 2006, Cyganowski, et al. 2008, Zhang \& Wang 2009, Maret et al. 2009, Ybarra \& Lada 2009, Ybarra et al. 2010, Dionatos et al. 2010, Barsony et al. 2010,  Takami et al. 2010); more important, however, is that for the first time sensitivity and spatial resolution have been achieved close to those reached from ground-based observations in the near-infrared, hence allowing H$_2$ observations in different spectral bands to be combined effectively.\\ 
We have started this kind of multiband analysis in a group of outflows from Class 0 sources observed with {\it Spitzer-IRS} between 5.5 and 37\,$\mu$m. The data have been presented by Neufeld et al. 2009 (hereafter Paper\,I), who gave an overview of the spectra along with the global physical properties {\it averaged over the entire emitting region}. The H$_2$ luminosities were
fitted by a NLTE model in which an admixture of gas temperatures is present within the spatial resolution element. A power-law distribution of the gas temperature is assumed, with the column of gas at $T$ and $T+dT$ proportional to $T^{-\beta} dT$; $\beta$ typically ranges between 2.3 and 3.3. Other main parameters of the model are the H$_2$ ortho-to-para ratio, that is found in all the flows apart from L1448 lower than the equilibrium value, and the particle density, which is typically lower than 10$^4$ cm$^{-3}$.
In a second paper, Nisini et al. 2010 (hereafter Paper\,II) have analyzed in deeper detail the outflow of L1157, deriving maps of the main parameters (H$_2$ column density, temperature, ortho-to-para ratio) that govern the shock physics at $\sim$ 2$^{\prime\prime}$ resolution, and giving detailed estimates of the molecular cooling. In particular, gradients of the H$_2$ column density are probed along the flow, with the highest values found in the blueshifted lobe, where $N(H_2)$ $\sim$ 3 10$^{20}$ cm$^{-2}$; the temperature spans from $\sim$ 250 to $\sim$ 1500 K, with the largest range of variations towards the intensity peaks close to the exciting source; the ortho-to-para ratio is significantly 
lower than the equilibrium value, especially in plateau regions between consecutive intensity peaks.\\
The present paper deals with three other outflows of our sample, namely  L1448, BHR71, and NGC2071, while a dedicated paper will be devoted to 
the analysis of VLA1623 (Liseau et al., in preparation). 
Together with L1157, they constitute a homogeneous sample of outflows from very young protostars.
Each object, however, has its own peculiarity: L1448 is a prototype of a source driving a molecular jet (Eisl\"offel 2000); BHR71 is one of the few outflows to be almost fully molecular (Giannini et al. 2004); NGC2071 is an example of a parsec-scale outflow of high mass (Stojimirovi\'c, Snell \& Narayanan 2008), and L1157 is maybe the most active outflow from the chemical point of view (Bachiller \& P\'erez Guti\'errez 1997). \\
The detailed description of morphology and physical characteristics of the three outflows discussed here have been given in Paper\,I and Melnick et al. 2008; here, we limit to briefly summarize some information useful for the comprehension of the present paper.\
The L1448 outflow (length $\sim$ 0.3 pc), originates from the Class\,0 source L1448-mm (L1448C, Bachiller, Andr\'e \& Cabrit 1991), subsequently identified in  {\it Spitzer} images as a binary system. At the tip of the northern lobe another triple system is located (Looney, Mundy \& Welch 2000), with a Class\,0 source, L1448-IRS3 (L1448-N) driving a compact molecular outflow.
Also the BHR71 outflow originates from an embedded binary protostellar system (IRS1 and IRS2, Bourke 2001), each driving a molecular outflow. IRS1, in particular, has been classified as a Class\,0 protostar (Bourke et al. 1997); the associated outflow extends in the north-south direction. The source IRS2 gives rise to a more compact outflow inclined of $\sim$ -36$^{\circ}$. Optical [SII] emission revealed HH associations (HH\,320 and HH\,321) along both the outflows (Corporon \& Reipurth 1997): these should be at very low degree of excitation, as demonstrated by the lack of emission both in H$\alpha$ and in near-infrared ionic lines (Giannini et al. 2004).
NGC2071 is a parsec scale bipolar outflow originating from a cluster of infrared sources. Among them, the source IRS1 has been initially suggested to be the outflow powering source, because it dominates the luminosity in the region 
(Persson et al. 1981); subsequently, however, images in H$_2$ emission (Aspin, Sandell \& Walther 1992) have demonstrated outflow activity around the source IRS3, that is also the major contributor to the luminosity at longer wavelengths (Walther, Geballe \& Robson 1991), likely due to its young age.\

The main aim of this work is to construct detailed maps of the physical parameters along each outflow. Our method is quite similar to that described in Paper\,II: this will allow us to directly compare our results with those obtained in L1157.
Such maps are of fundamental importance for elucidating, albeit in a small sample, the typical physical conditions of molecular outflows, in terms of temperature, density, and mass of the shocked gas. Our results also put observational constraints on models for partially (or non)-dissociative shocks. We also note that the objects presented here are going to be observed in the far-infrared with the {\it Herschel} telescope as part 
of both Guaranteed and Open Time programs. In particular, they belong to the sample of the {\it WISH} program, devoted to the understanding of the chemistry of shocks in protostellar outflows: the maps and the physical parameters given in the present paper will be of fundamental importance in deriving the water abundance along the flow, for example.\\
The paper is organised as follows: in \textsection\ref{sec:sec2} we briefly describe the observations and the results obtained; \textsection\ref{sec:sec3} is dedicated to the analysis of the H$_2$ emission, in which we have adopted both an LTE and  non-LTE approximation for the line analysis, while in \textsection\ref{sec:sec4} we analyze a few spots where HD lines were detected. \textsection\ref{sec:sec5} and \textsection\ref{sec:sec6} are dedicated to a brief comparison with studies of Paper\,I/II and a discussion on the outflows' energetics; finally, in \textsection\ref{sec:sec7} we summarize our conclusions.

\begin{figure*}
\includegraphics[angle=-90,scale=0.7]{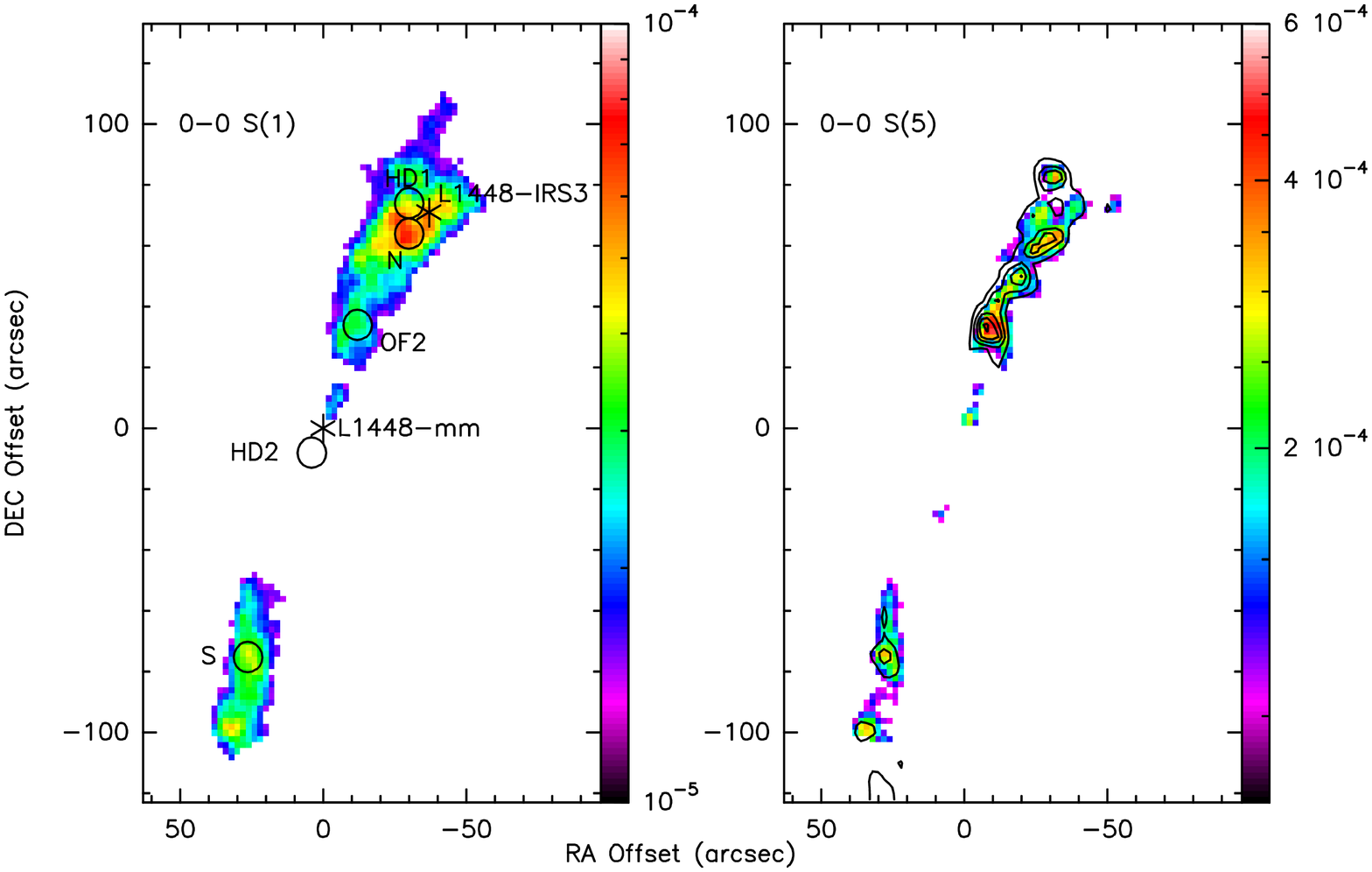}
\caption{L1448 maps of the 0-0 S(1) and 0-0 S(5) lines, this latter with overlaid contours
of the 1-0 S(1) line in steps of 20\% of the peak, which is 5.0 10$^{-5}$ erg s$^{-1}$ cm$^{-2}$ sr$^{-1}$.
Vertical bars indicate the 0-0 lines brightness in the same units. Regions labelled HD1 and HD2 represent peaks of HD
emission, where the R(3) and R(4) are detected, while regions N and S are peaks of H$_2$ emission in the northern and southern lobe, 
respectively.Following the nomenclature of Dionatos et sl. (2009), we have also marked the 'OF2' region, where other near-infrared 
H$_2$ lines have been detected.} 
\label{L1448_images}
\end{figure*}

\begin{figure*}
\includegraphics[angle=-90,scale=0.7]{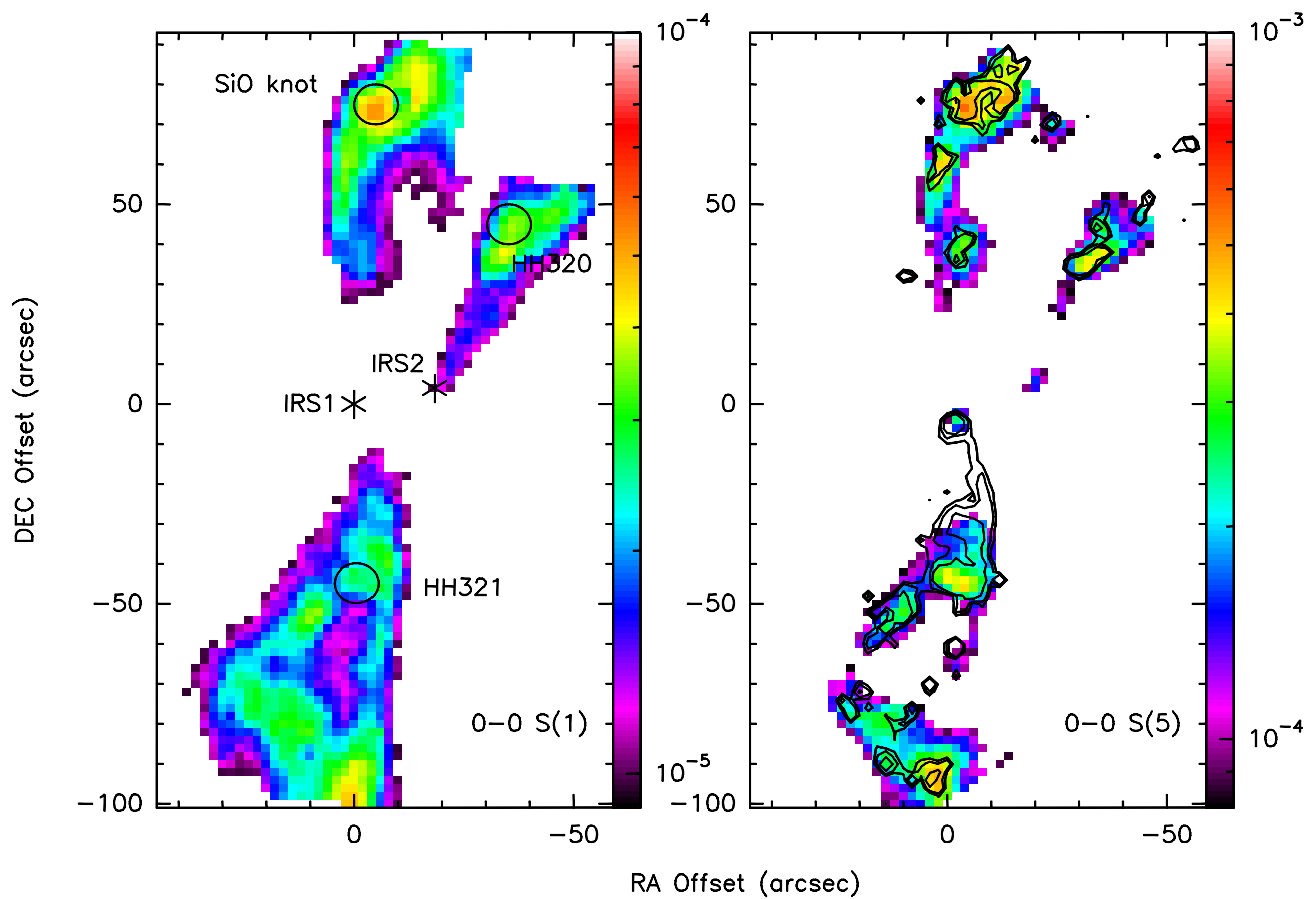}
\caption{As for in Figure \ref{L1448_images} for BHR71. The peak of the 1-0 S(1) line  is 
1.8 10$^{-4}$ erg s$^{-1}$ cm$^{-2}$ sr$^{-1}$. The region south IRS1, where the 0-0S(5) is not detected (in contrast with the 1-0S(1) line)  was contaminated by strong stripes and artifacts, that have been removed within the data reduction procedure. The positions of two Herbig-Haro objects (HH 320/HH 321) and that of a peak of SiO emission (Gusdorf et al. 2011), are indicated.}
\label{BHR71_images}
\end{figure*}

\begin{figure*}
\includegraphics[angle=-90,scale=0.7]{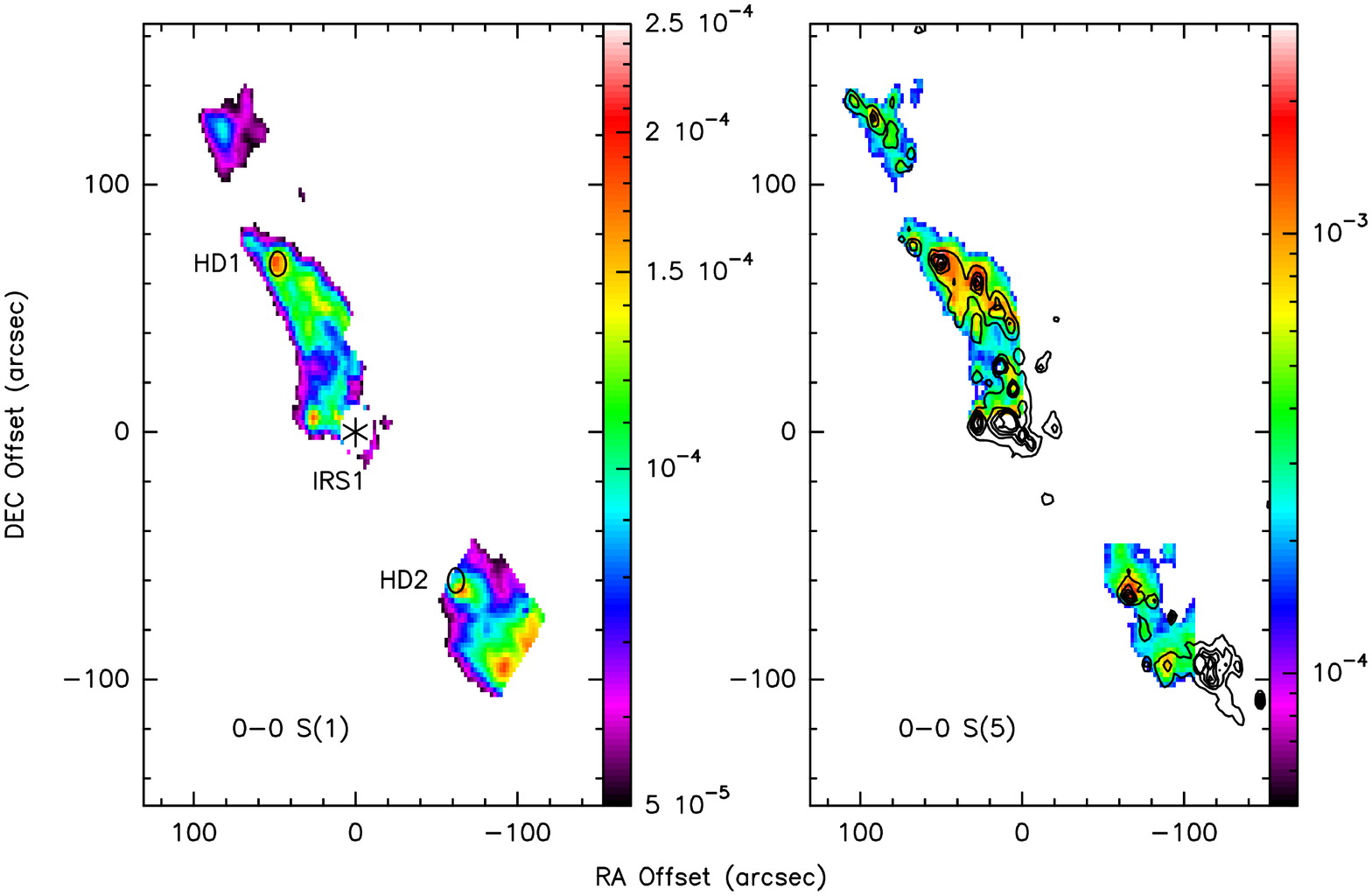}
\caption{As for in Figure  \ref{L1448_images}  for NGC2071. The peak of the 1-0 S(1) line  is 
9.0 10$^{-4}$ erg s$^{-1}$ cm$^{-2}$ sr$^{-1}$. The central region close to IRS 1 results saturated in the {\it IRS} map and the very southern 
bow-shock was not mapped. Regions labelled HD1 and HD2 represent peaks of HD
emission, where the R(3) and R(4) are detected.}
\label{N2071_images}
\end{figure*}
\section{Observations and results\label{sec:sec2}} 
All the three outflows have been observed during the Cycle 4 of the {\it Spitzer} mission (February-March 2008), with the
exception of a single 1$^{\prime} \times 1^{\prime}$ region mapped previously toward each of BHR71 and NGC2071. 
{\it IRS} observations in the full spectral range (5.2-36.5 $\mu$m) were carried out, with the Long-High (LH), Short-High (SH)
($\lambda/\Delta\lambda \sim$ 600) and Short-Low ($\lambda/\Delta\lambda$ between 64 and 128) modules. The total integration 
time in each object is between 19 and 23 hours.  
Each outflow map has been obtained by combining several 1$^{\prime} \times 1^{\prime}$ maps, which were arranged along the outflow 
axis by stepping the {\it IRS} slit by one-half of its width in the direction perpendicular to its length. In the SH and LH modules, the slit was also stepped parallel to its length by 4/5 (SH) and 1/5 (LH). Further details on the observational strategy along with the data reduction procedure are given in Paper\,I. \\
The molecular emission is largely dominated by H$_2$ emission that has been detected in all of three outflows.
All the pure rotational lines of the ground vibrational level accessible with {\it Spitzer}, i.e. from S(0) to S(7), have been detected. 
In addition, a few water lines (whose analysis will be the subject of a future paper), and the HD R(3) and R(4) rotational lines have been detected in a few spots along the L1448 and NGC2071 outflows. 
In Table 1 we report the H$_2$ and HD brightness in selected areas (of radius 5$^{\prime\prime}$) that represent 
peaks of emission or mid-infrared counterparts of peaks at other wavelengths.  
In the same table, we also give the brightness of the 1-0\,S(1) line at 2.12\,$\mu$m computed for the same area. The 1-0\,S(1) image of BHR71
was retrieved from the ESO archive\footnote{available at http://archive.eso.org/eso/eso-archive-main.html}, while the other two were taken from Davis \& Smith (1995) and Eisl\"offel (2000). The images were first rebinned at the {\it IRS} spatial sampling (of $\sim$ 2$^{\prime\prime}$/px) and then flux-calibrated by using bright field stars whose photometry was retrieved from the
2MASS catalog. \\
Figures \ref{L1448_images}, \ref{BHR71_images} and \ref{N2071_images} show the images of the 0-0 S(1) and 0-0 S(5) lines, overlaid with contours of the 2.12\,$\mu$m line intensity in
steps of 20\% from the peak. The overall morphology of the 0-0 S(1) and 0-0 S(5) lines is similar; 
this latter, in particular, correlates well with the peaks of the 1-0 S(1) line (right panels), given the quite
similar excitation energy, that also makes them probes of the same gas component at thousands of Kelvin.\\
Figure \ref{L1448_images} shows the line maps of L1448, where we have labelled the peaks of the H$_2$ emission along with the positions of the two HD peaks that are revealed by the R(3) and R(4) lines. We have identified two peaks (N and S), in the northern and in the southern lobe, respectively; peak N, in particular, is few arcseconds south-east of the source L1448-IRS3. Following the nomenclature of Dionatos et al. (2009), we have also labelled the 'OF2' position, for which near-infrared spectra have been obtained, and that we will discuss in greater detail in \textsection\ref{sec:sec3.2.2}. 

In Figure \ref{BHR71_images}, we have labelled the mid-infrared counterparts of the 
two Herbig-Haro objects in BHR71 (HH\,320 and HH\,321), firstly seen by Bourke (2001) at 2.12\,$\mu$m and then observed
spectroscopically in the near-infrared by Giannini et al. (2004). In the same image, we
have also identified the mid-IR counterpart of a SiO peak located in the northern lobe of the outflow excited by the source IRS1, which
has been recently investigated by Gusdorf et al. (2011). In agreement with that paper, we label this peak as 'SiO knot'.\\
Finally, Figure \ref{N2071_images} represents the portion of the outflow of NGC2071 mapped with {\it IRS}. The overall morphology of the S(5) line follows that of the 1-0\,S(1) line, although the {\it IRS} maps do not cover the very southern region and are saturated around the position of the source IRS1. We detected two spots of HD emission, one in each outflow lobe, that are also correlated with the H$_2$ peaks.

\section{Analysis of the H$_2$ emission\label{sec:sec3}} 

\subsection{Maps of averaged parameters\label{sec:sec3.1}} 
In this section, we describe the analysis of the H$_2$ lines, used to derive the maps of the main physical parameters characterizing the molecular gas. As a first step, we will analyse the pure rotational lines only: given their relatively low excitation energy, most of them are  observed both along the outflow axis and in the external parts where the gas interacts with the environment: consequently, we will be able to probe portions of the outflows where gradients in the parameters are expected to occur. As a first step, we assume LTE conditions for the 0-0 lines. This assumption is supported by the low values of the critical densities, that never exceed 5 10$^6$ cm$^{-3}$ for $T$ between 500 and 2000 K (Le Bourlot, Pineau des For\^ets \& Flower 1999). For the analysis, we consider only collisions with H$_2$ molecules, since the critical densities do not substantially change if we include also collisions with H and He. Possible departures from LTE will be considered further on (see \textsection\ref{sec:sec3}.2). Our method follows that of Paper\,I and Paper\,II: for each pixel we have constructed the rotational diagram\footnote{This is the plot ln(N$_{vJ}$/g$_{J}$g$_{s}$) vs. E$_{vJ}$, N$_{vJ}$ being the column density of the level with vibrational state $v$ and rotational state ${J}$, $s$ the spin quantum number, ${g}$ the degeneracy, and E$_{vJ}$ the excitation energy of the upper level (in K).} from the S(0) to S(7) lines, plotting the column density in each rotational state as a function of the excitation energy.
Such a diagram resembles a straight line if all the emitting gas is in LTE at the same temperature (e.g. Gredel 1994), while it follows a curved distribution in the more general case where gas at different temperatures are within a spatial resolution element and/or are intercepted along the line of sight. 
Since we are observing post-shocked regions, where the temperature ranges from thousands to hundreds of Kelvin, 
we have fitted the observations
by assuming a power law temperature distribution of the form  {\it dN $\propto$ T$^{-\beta}$ dT}. Shock models predict  $\beta$ typically $\ga$ 3.8 in bow C-type shocks (Neufeld \& Yuan 2008), while lower values are expected in J-type shocks, where the shock conditions are strong enough to excite lines at high excitation energy (up to $\sim$ 30\,000 K, e.g. Flower et al. 2003). Consequently, we have taken $\beta$ as a free parameter. It is important to underline that also a change in the H$_2$ density influences the curvature shape in the rotational diagram due to the different critical densities of the lines. This implies that a certain degree of degeneracy exists between  $\beta$ and  {\it n(H$_2$)}, and that, for the LTE conditions we are assuming, 
the $\beta$ values we fit have to be considered as upper limits.\\

To derive the H$_2$ column density, we have integrated the power law between a minimum (T$_{min}$) and maximum (T$_{max}$) value
of temperature. Since we expect that temperatures at thousands of Kelvin contribute to the column density of the pure rotational lines only marginally, we have fixed T$_{max}$ at 4000 K (and we have verified  {\it a posteriori} that 
{\it N(H$_2$)} varies of only few percent if T$_{max}$ = 5000 K). Conversely, since we expect that the bulk of the emission traced by the observed lines comes from cold gas, we have chosen  not to take T$_{min}$ as constant over the entire flow; rather we adopted for each pixel as T$_{min}$ the temperature fitted in the rotational diagram  of the S(0) and S(1) lines only, i.e. the coldest gas component we are able to trace. T$_{min}$ ranges typically between $\sim$ 150 and 400 K.\\

Another free parameter in our fit is the ortho-to-para (OPR) ratio value, which can significantly depart from the equilibrium value of 3
at the low temperatures probed by the 0-0 lines. Non-equilibrium values are evident in a 'zigzag' behaviour of the rotational diagram (see Figure\,16 of Paper\,I), in which the states  with even-$J$ lines lie systematically above those with odd-$J$. The departure from the equilibrium value
decreases gradually with increasing $J$, such that the highest energy levels (corresponding to S(6) and S(7) lines) have an OPR $\sim$ 3; we have however considered a single value for all the lines in order to minimize the number of free parameters. This assumption thus implies that the OPR we derive is indeed an average value. \\

Finally, the other parameter affecting the observed line intensities, and consequently the column densities, is the visual extinction. Since this affects very marginally the fluxes of the 0-0 lines, we have taken A$_V$ constant toward each shocked region. The extinction values have been taken from literature: we have assumed A$_V$=2 mag for BHR71 (Giannini et al. 2004), A$_V$=6 mag for L1448 (Nisini et al. 2000), and  A$_V$=13 mag for NGC2071 (Melnick et al. 2008). In the range from 1 to 13 $\mu$m we adopted the extinction law of Rieke \& Lebofsky (1985) and that of  Mathis (1998) from 13 to 23 $\mu$m, which was extrapolated at 28 $\mu$m. \\

\begin{figure}
\includegraphics[angle=-90,scale=0.65]{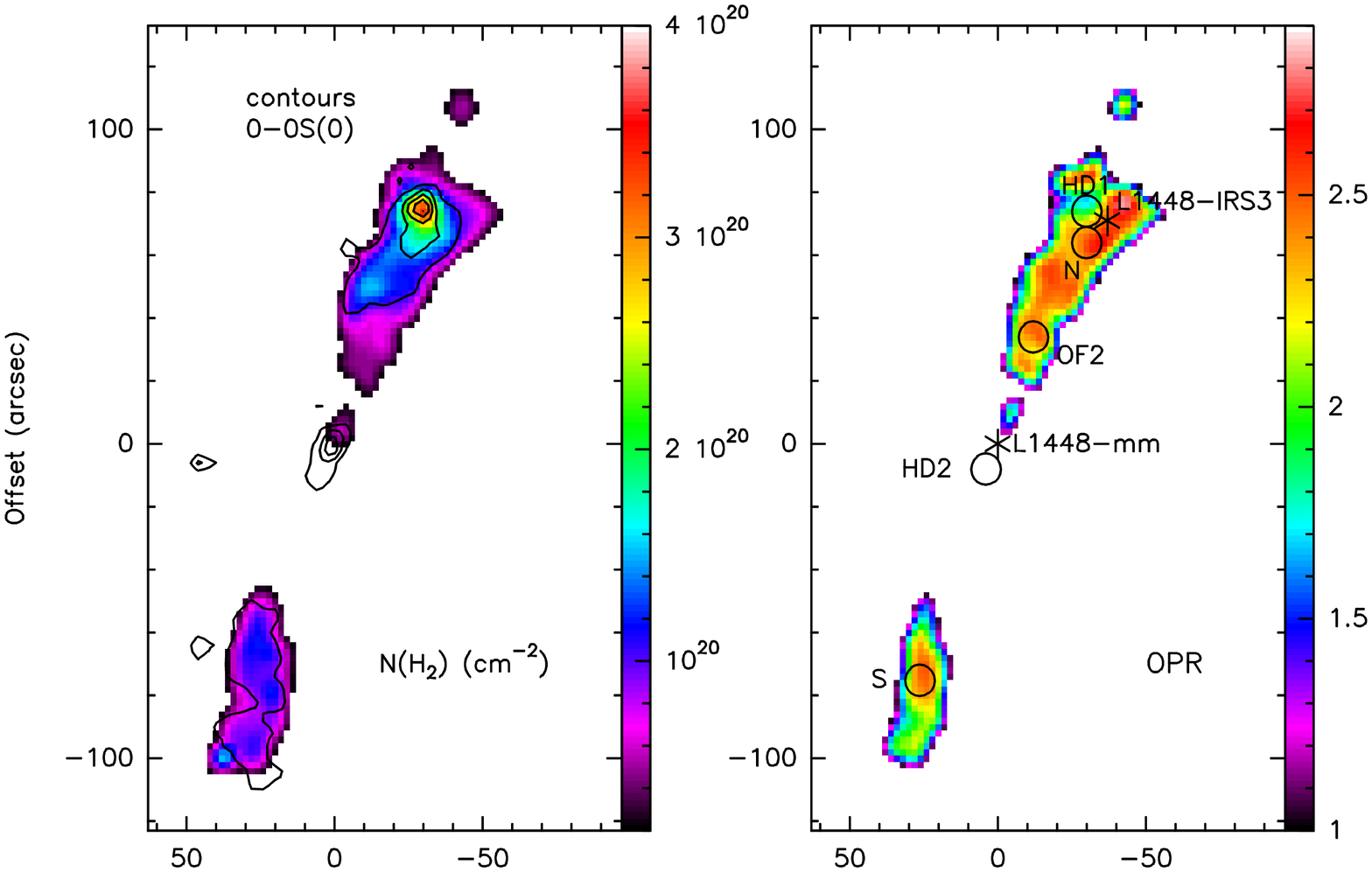}
\includegraphics[angle=-90,scale=0.65]{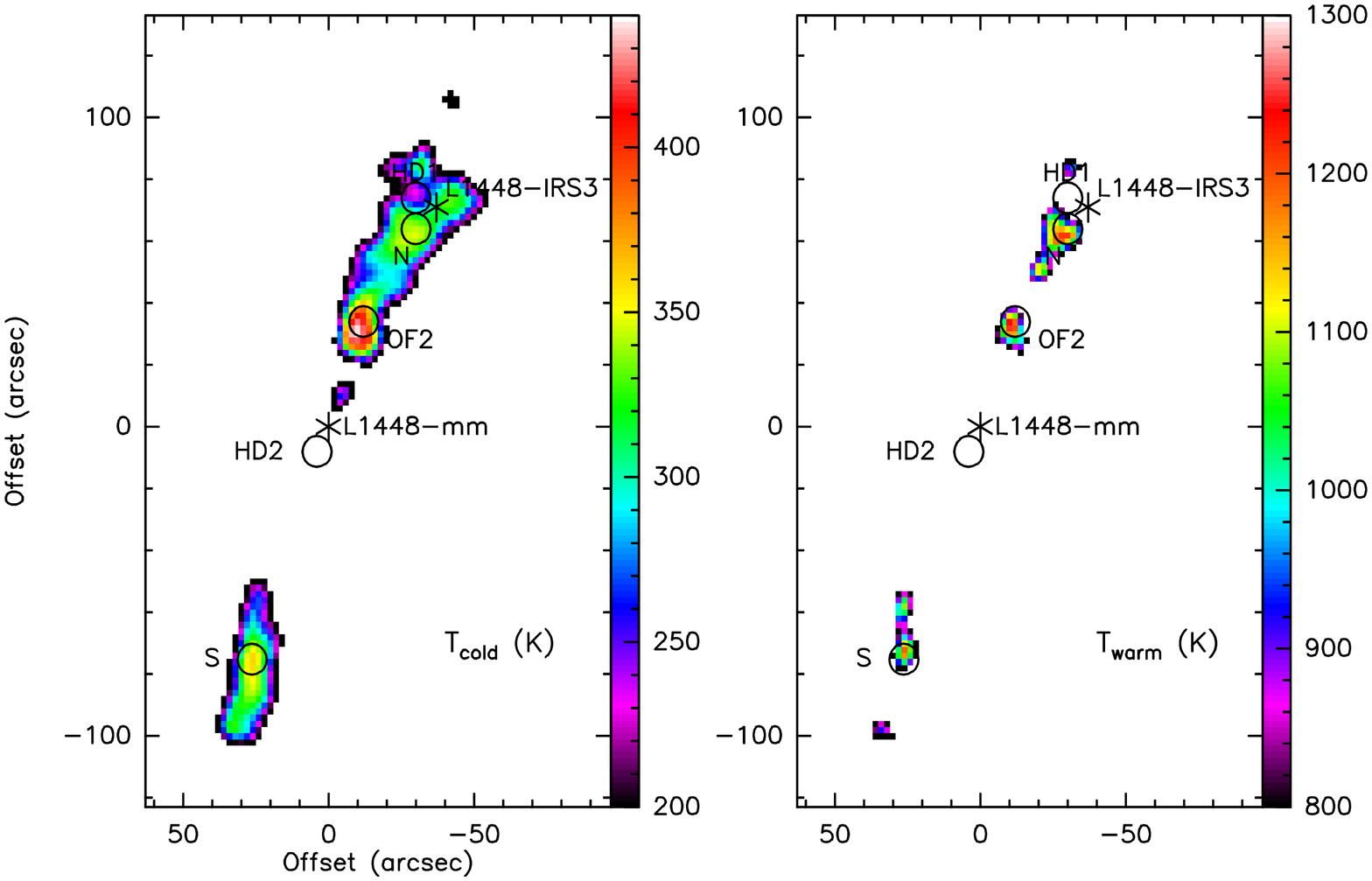}
\caption{Map parameters of L1448. Top panels: total H$_2$ column density (left); ortho-to-para ratio (right); bottom panels: 'cold' (left) and 'warm' (right) gas temperature
components, respectively derived from the fit through S(0)-S(1)-S(2) and S(5)-S(6)-S(7) lines.
In the top left panel contours of brightness of the S(0) line are shown for comparison, plotted in steps of 20\% from the peak.
Vertical bars indicate the values of the parameters in the units given in parentheses.}
\label{L1448_all}
\end{figure}

\begin{figure}
\includegraphics[angle=-90,scale=0.65]{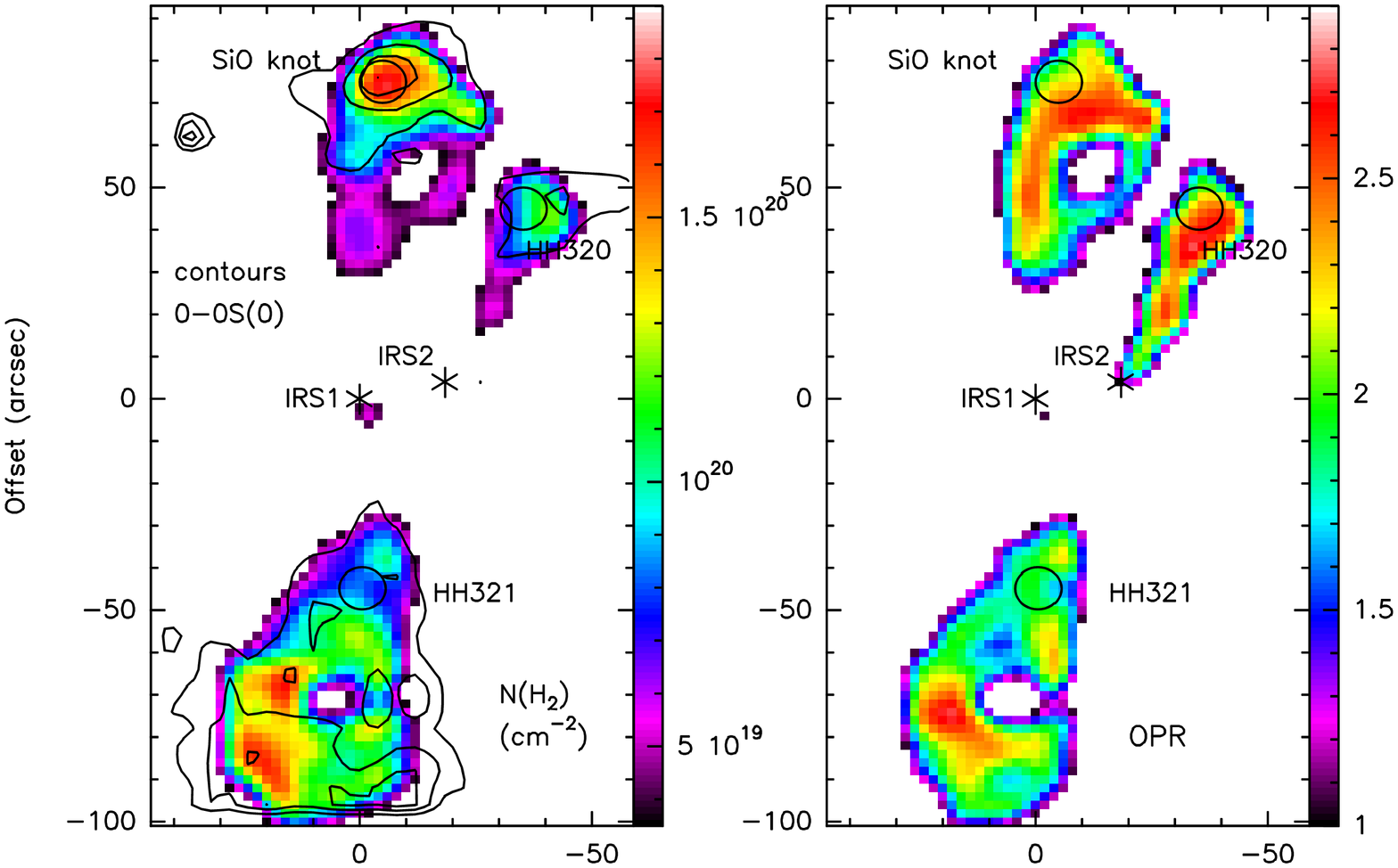}
\includegraphics[angle=-90,scale=0.65]{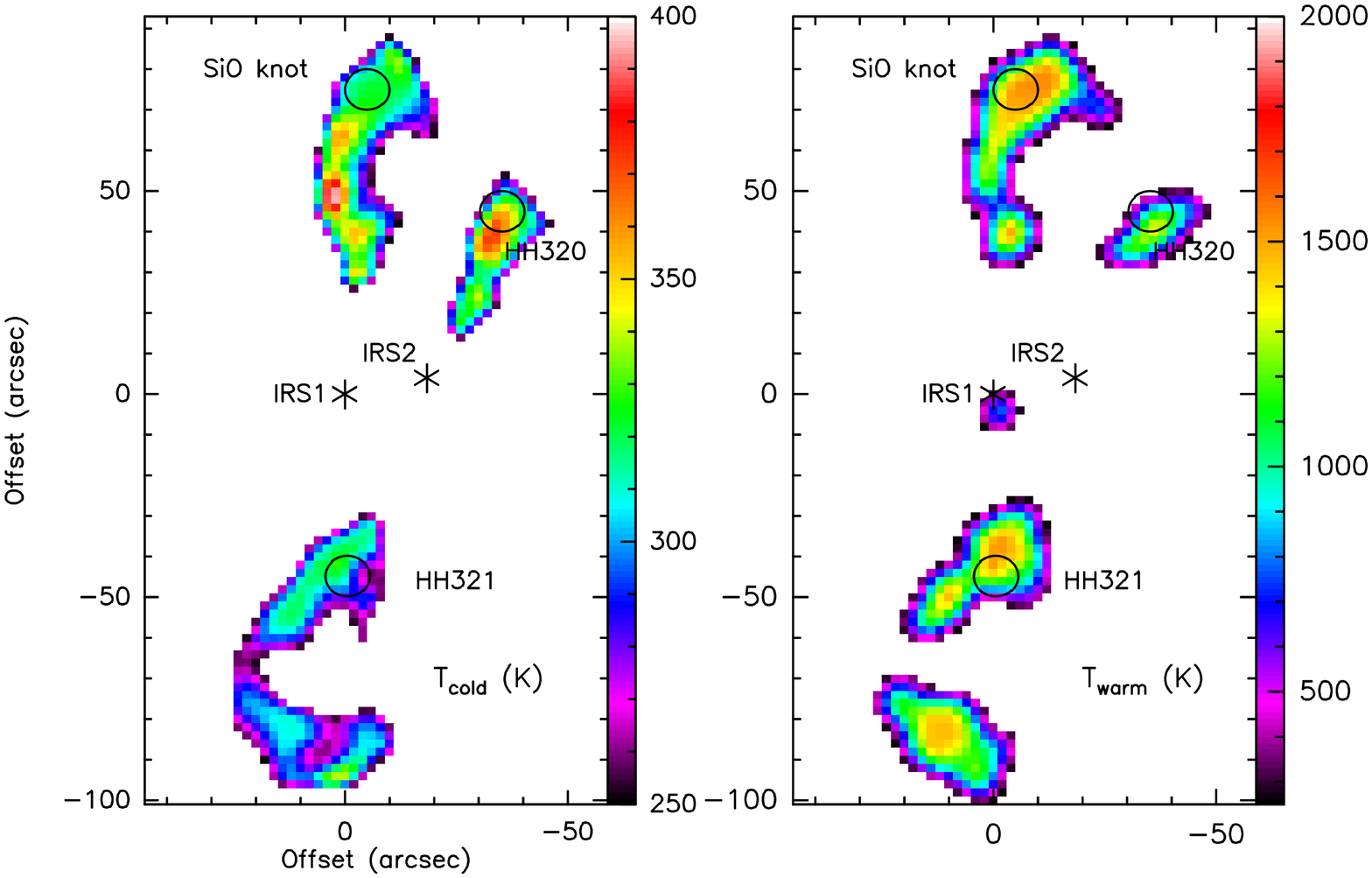}
\caption{As in Figure \ref{L1448_all} for BHR71.}
\label{BHR71_all}
\end{figure}

\begin{figure}
\includegraphics[angle=-90,scale=0.65]{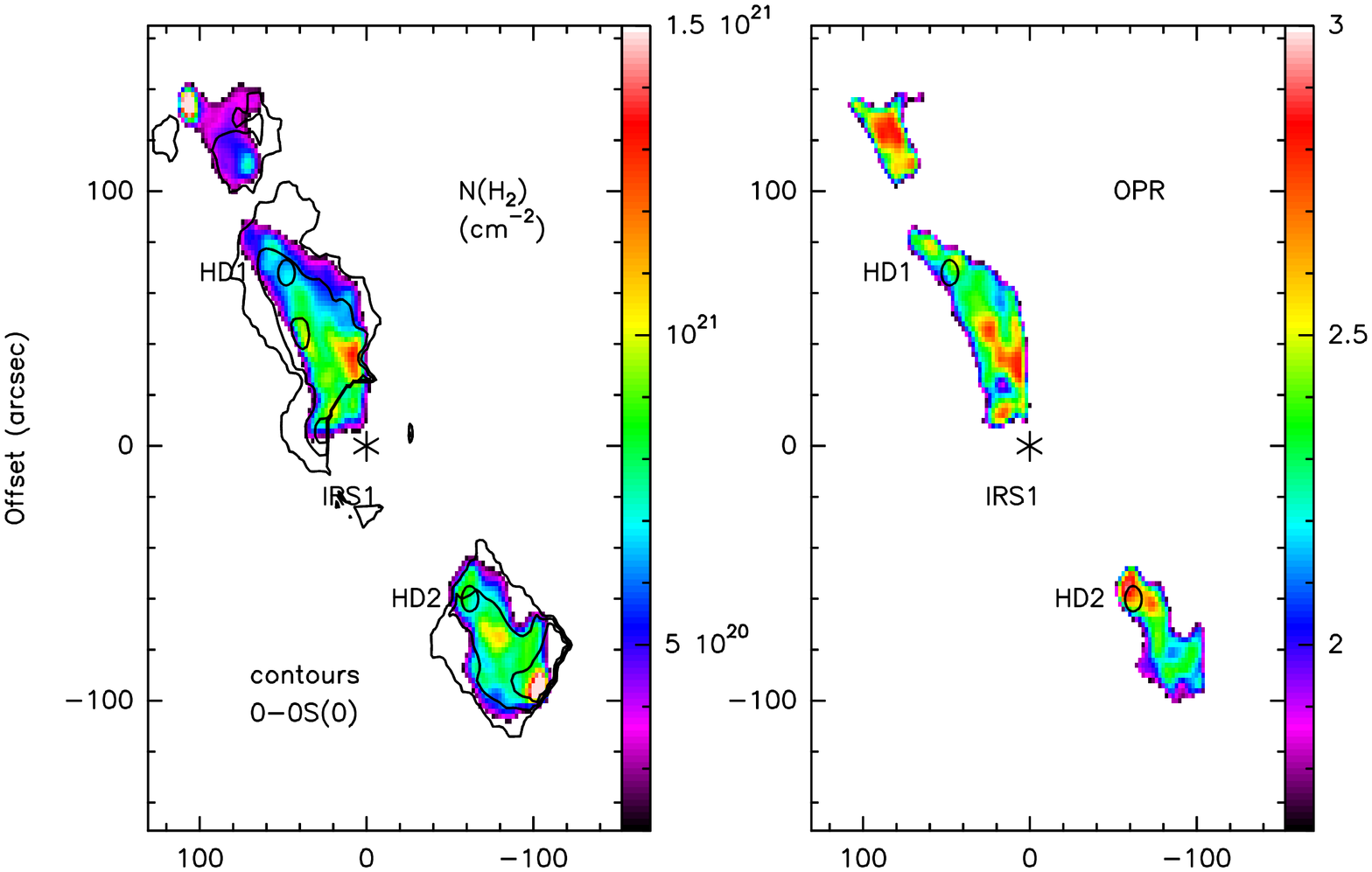}
\includegraphics[angle=-90,scale=0.65]{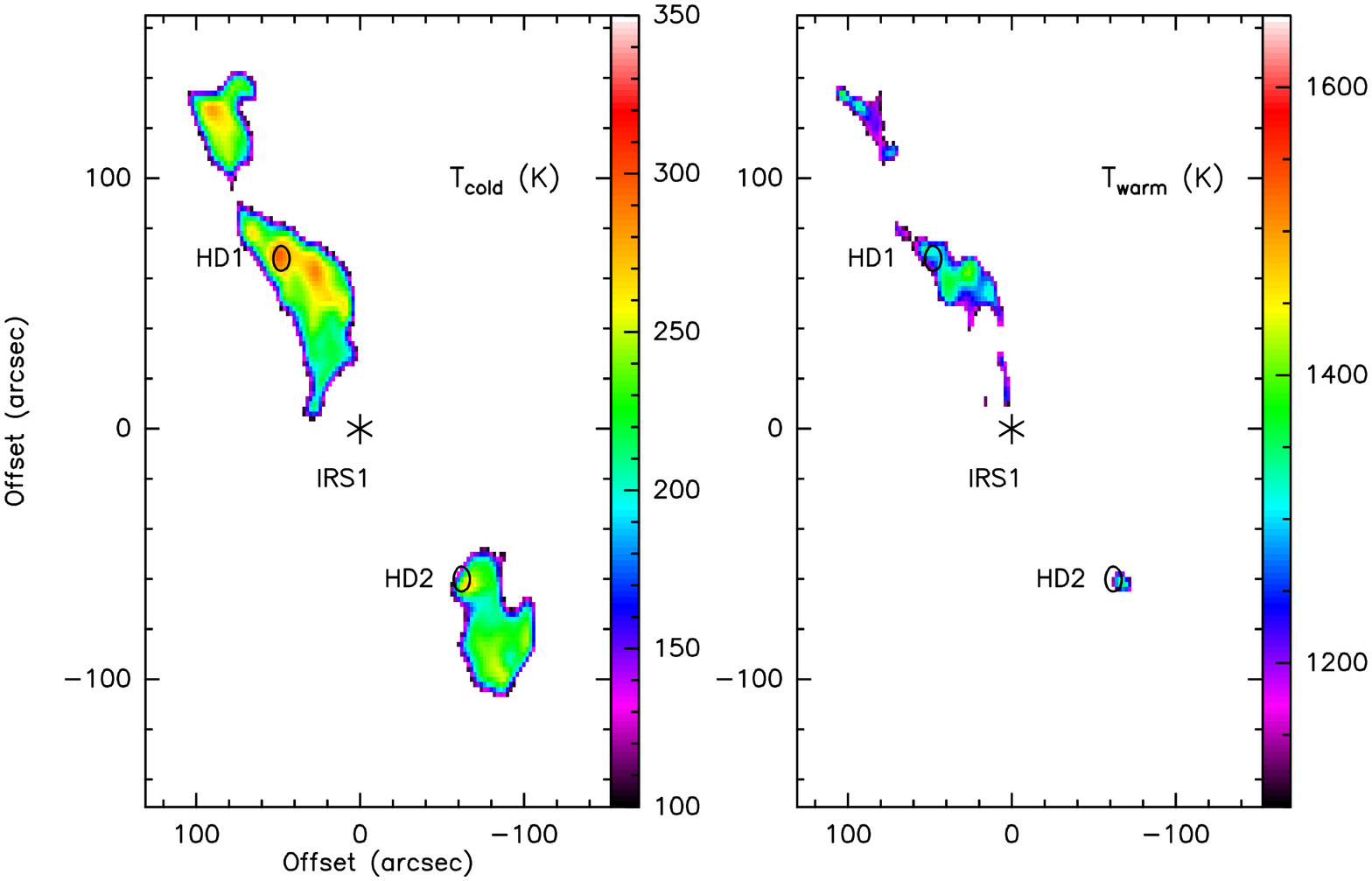}
\caption{As in Figure \ref{L1448_all} for NGC2071.} 
\label{N2071_all}
\end{figure}

Summarizing, our fits have three free parameters: index $\beta$ of the {\it dN} vs. {\it T} power law, the OPR, and the total  column density
{\it N(H$_2$)}. These
three parameters have been fitted for each pixel where at least four lines are detected at S/N $\ge$ 3 by means of a $\chi^2$ minimization. Typical uncertainties in the parameters are as follows: {\it N(H$_2$)}: $\sim$ 10\%, OPR: $\pm$ 0.2, $\beta$: $\pm$ 0.2. Together with the maps of the above parameters, we also derived two maps of temperature, that represent an estimate of the maximum (T$_{warm}$) and the minimum (T$_{cold}$) temperatures contributing to the observed emission; these have been  obtained as fits of the S(0)-S(1)-S(2), and the S(5)-S(6)-S(7) lines, respectively, once corrected for the value of the OPR in the considered pixel. \\
In Table \ref{tab:tab2} we give the value of each parameter averaged  over the entire map, with, in parentheses, the minimum and maximum value attained at particular positions along the flow. At this stage we do not give the fitted values of $\beta$, which will be derived again in NLTE approximation, i.e. once the degeneracy with {\it  n(H$_2$)} has been removed, see \textsection\ref{sec:sec3.2.1}.\\ 
From an inspection of Table 2, it is clear that all the parameters vary significantly, and that 
similarities and differences are apparent among the objects. These are discussed below with reference to the individual maps
(Figures \ref{L1448_all} - \ref{N2071_all}). In the top left panel we present the H$_2$ total column density: 
a good correlation between {\it N(H$_2$)} and the intensity peaks in the S(0) line (whose contours are overlaid) is clearly evident, implying that the large majority of the molecular gas is at low temperature. Variations up a factor $\sim$ 4-7 are evident in all the three flows, with {\it N(H$_2$)} 
ranging between 0.4-4 10$^{20}$ cm$^{-2}$ in L1448, 3-20 10$^{19}$ cm$^{-2}$ in BHR71, and 5-18 10$^{20}$ cm$^{-2}$ in NGC2071. In L1448
(Fig.\ref{L1448_all}), the maximum {\it N(H$_2$)} is attained close to the apex of the northern bow shock: this is the region where the gas is strongly compressed because of the interaction of the high-velocity outflowing gas with both the high-density clump hosting the system of L1448-IRS3, and the associated outflow; more constant values occur in the southern, red-shifted lobe. We do not exclude, however, the possibility that a significant column density variation can occur at the apex of the bow, which was not observed with {\it IRS}.  
A similar behaviour is recognizable in the {\it N(H$_2$)} column density map of BHR71 (Fig.\ref{BHR71_all}), where peaks of more than 10$^{20}$ cm$^{-2}$ are found at the apex of the two outflows driven by sources IRS1 and IRS2. These are the regions where stronger shock conditions
are evident both at optical-infrared wavelengths (HH\,320/321) and in the sub-mm (SiO knot).
A more constant behaviour is found in the {\it N(H$_2$)} values fitted in the NGC2071 outflow (Fig.\ref{N2071_all}). Both lobes present peaks
of {\it N(H$_2$)} $\sim$ 10$^{21}$ cm$^{-2}$, particularly in the parts closest to the cluster of young sources at the centre of the two lobes. \\
The OPR, shown in the top right panels of Figures \ref{L1448_all}-\ref{N2071_all}, presents significant departures from the equilibrium value of 3 in all three outflows. Similar values have been found, e.g., on NGC1333 (Maret et al. 2009),  on the Wasp-Waist nebula (Barsony et al. 2010), and
on L1157 (Paper\,II). In particular, low values of OPR, typically between 2.0 and 2.5 are often found along the external parts of the shock, likely because these regions are  where the warm gas in the jet  entrains the cold one of the ambient medium. 
Conversely, the highest values are generally found where gas components up to $\sim$ 2000\,K are probed (see below); we note,
however, that the equilibrium value of 3 is never reached: this indicates that the shock activity, presently on-going, heats the gas for timescales shorter than those required for a complete para-to-ortho conversion.  
\\
In the bottom right panels of Figs. \ref{L1448_all}-\ref{N2071_all} we show the maps of T$_{warm}$. As anticipated above, we derive this quantity from the linear fit through the most excited 0-0 lines, i.e. S(5)-S(6)-S(7). These are detected in L1448 and in NGC2071 only close to the jet axis, while in BHR71 the high-$J$ emission appears more extended. T$_{warm}$ ranges typically from 1000 to 2000 K, with lower and almost constant values found along the jet axis of L1448 and NGC2071; in BHR71 peaks of temperatures are found at the locations of the HH objects and close to the SiO knot. The typical difference between T$_{warm}$ and T$_{cold}$ (see bottom left panels), which gives an idea of the temperature gradient, is between 900 and 1200 K: this indicates that these 'warm' regions are also dynamically active. Of course, we cannot derive a similar conclusion on the regions where the high-$J$ lines remain undetected; however we note that these are the same where also T$_{cold}$ is low (typically $\sim$ 200-250\,K): hence it is reasonable that, even if a gradient up $\sim$ 1000\,K should exist, the maximum temperature would not be high enough  to excite bright high-$J$ lines.

\input{tab1.tex}

\input{tab2.tex}

\subsection{NLTE analysis\label{sec:sec3.2}} 
In this section, we will relax the assumption of the LTE, valid for the pure rotational lines, by including in our
analysis near-IR rovibrational lines coming from levels with critical density significantly higher than the $v$=0, J$\le$ 9 levels. 
In particular, maps of the 2.12\,$\mu$m lines will be used to infer on the H$_2$ density (\textsection\ref{sec:sec3}.2.1), while spectroscopic observations performed in selected regions will be used to construct rotational diagrams covering a large range of excitation energies (\textsection\ref{sec:sec3}.2.2).

\subsubsection{NLTE maps\label{sec:sec3.2.1}} 
\begin{figure}
\includegraphics[angle=-90,scale=0.7]{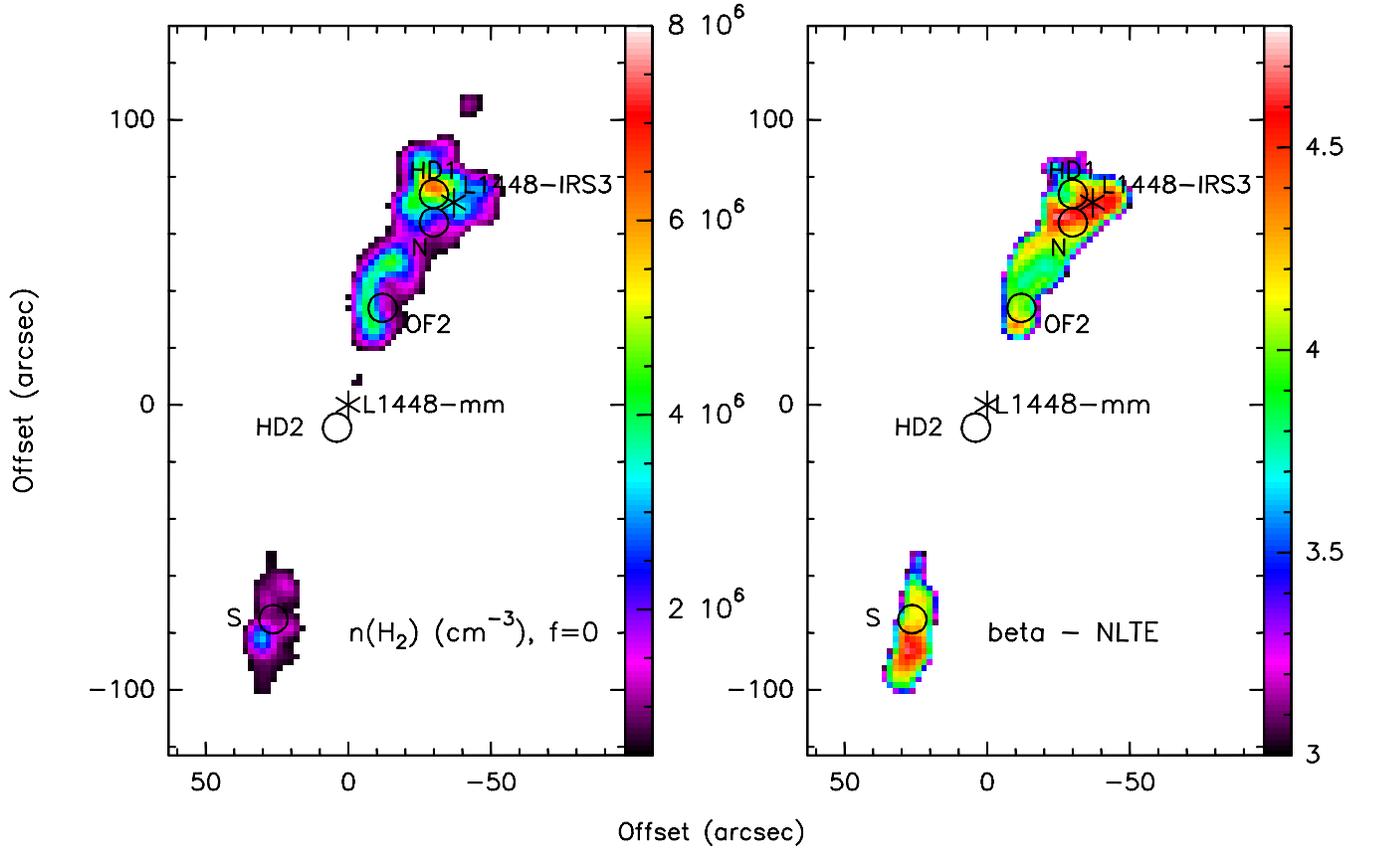}
\caption{Map parameters obtained under NLTE in L1448. Left: map of  H$_2$ particle density; no collisions with atomic hydrogen are considered; right: map of the power low index $\beta$ of the relationship {\it dN $\propto$ T$^{-\beta}$ dT}.}
\label{L1448_NLTE}
\end{figure}

\begin{figure}
\includegraphics[angle=-90,scale=0.7]{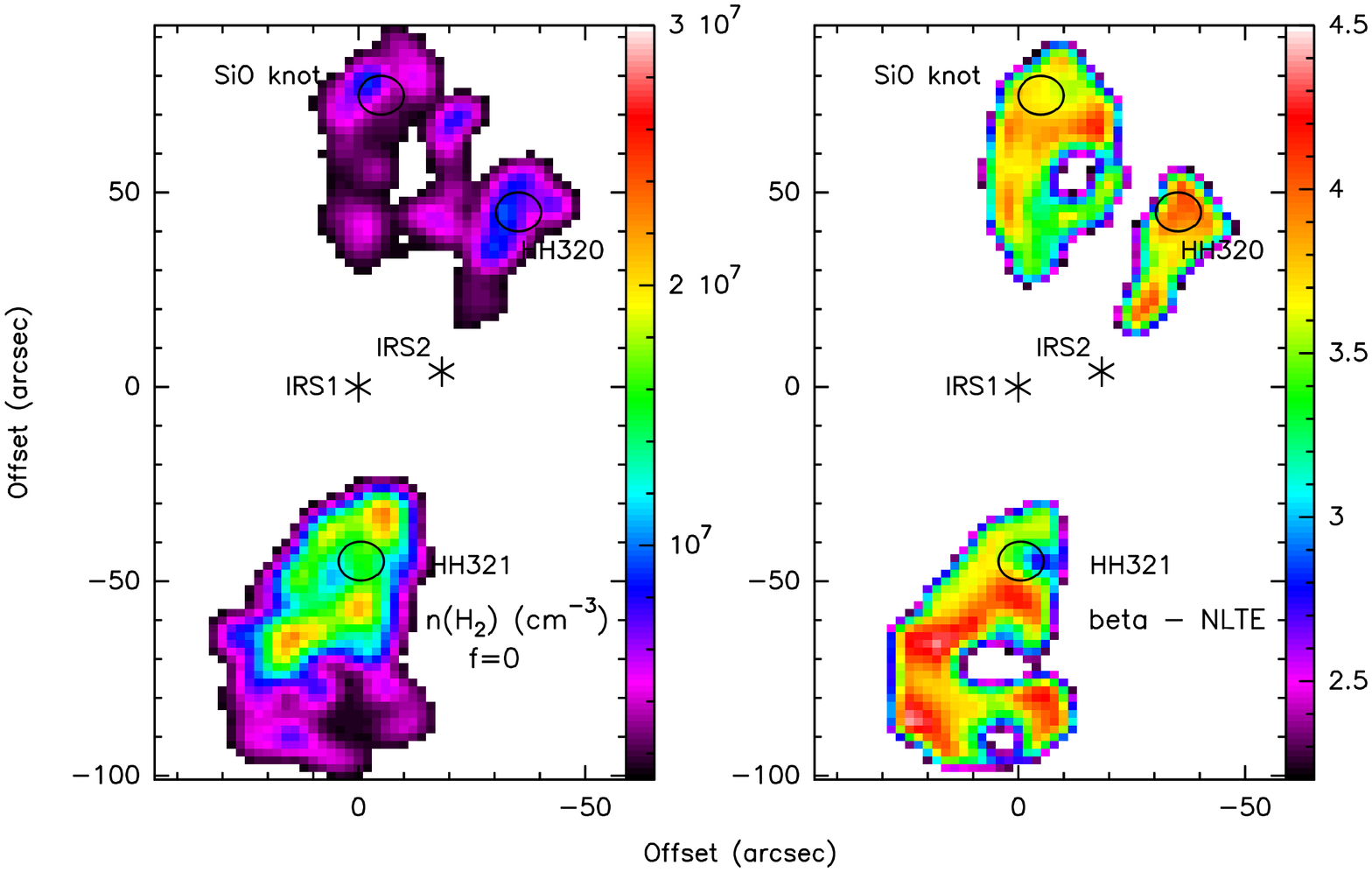}
\caption{As in Figure \ref{L1448_NLTE} for BHR71.}
\label{BHR71_NLTE}
\end{figure}

\begin{figure}
\includegraphics[angle=-90,scale=0.7]{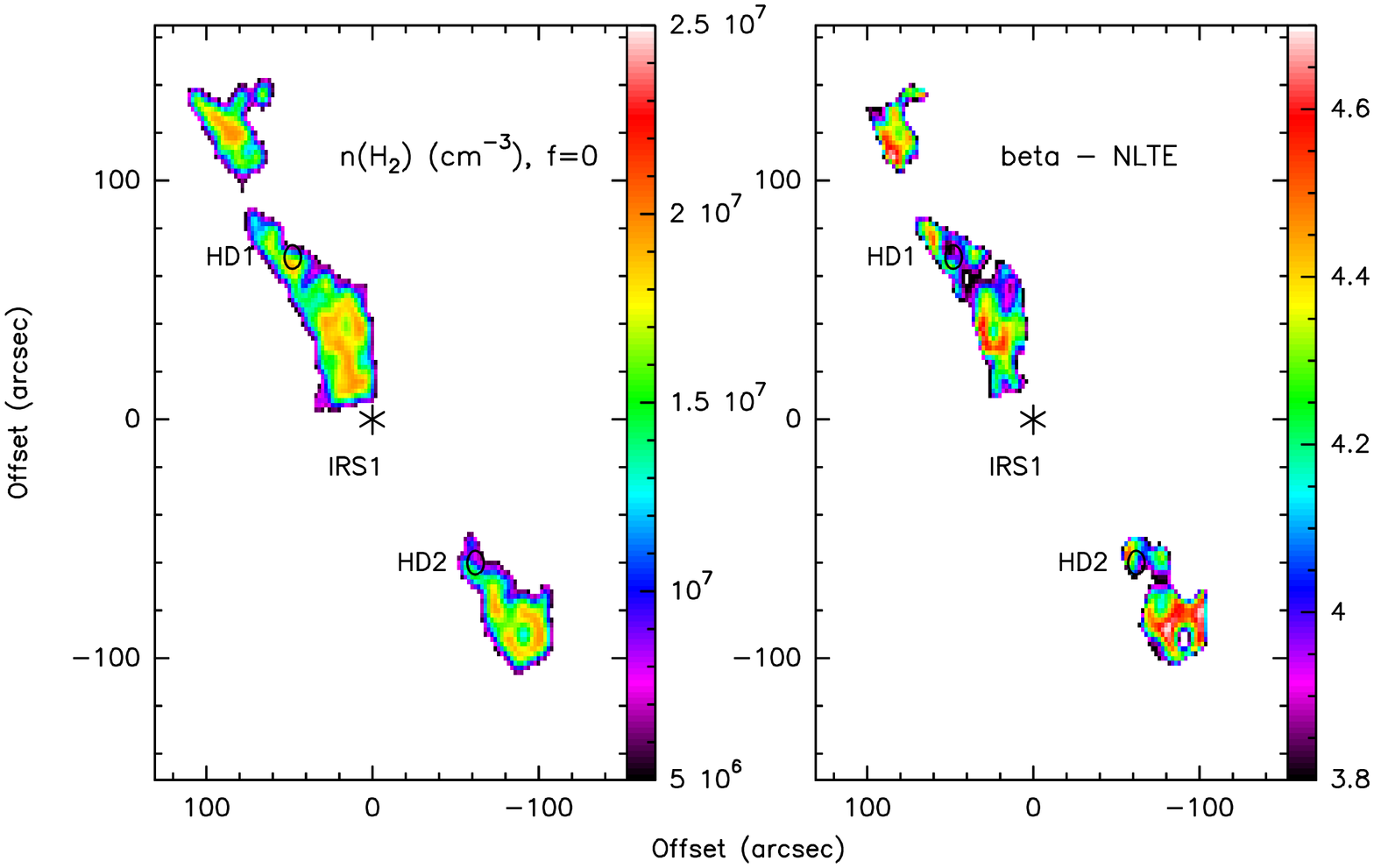}
\caption{As in Figure \ref{L1448_NLTE} for NGC2071.}
\label{N2071_NLTE}
\end{figure}

As anticipated above, we will exploit the high critical density of the 2.12\,$\mu$m line (ranging from 1.8 10$^{10}$ and 4.2 10$^{8}$ cm$^{-3}$ for 1000 K $\le$ T $\le$ 2000 K and collisions with H$_2$ only), to fit the lines under NLTE conditions and to derive 
a map of the H$_2$  particle density. Moreover, since we are deriving  {\it n(H$_2$)} independently from the  {\it Spitzer} lines, we also expect to remove the degeneracy between {\it n(H$_2$)} and  $\beta$ in the rotational diagrams. We stress that all these parameters (as shock velocities, temperature, and ionization), are expected to change rapidly on sub-arcsecond scales, ie. much smaller than our spatial resolution. In this sense,
the densities and $\beta$ values derived here are therefore mean values.
Our NLTE model consists of 24 and 26 levels for ortho- and para-H$_2$, respectively. We have considered collisions with H$_2$ and atomic hydrogen, 
whose de-excitation coefficients are taken from Le Bourlot, Pineau des For\^ets \& Flower (1999). 
We have taken as input parameters the values of {\it N(H$_2$)}, T$_{min}$, and OPR fitted for each pixel in LTE approximation. Indeed, the inclusion
in the fit of the 1-0\,S(1) line does not affect either the first two parameters, which depend mainly on the column density of the 0-0 lines
 at low-$J$, or the value of the OPR, which was obtained as an average over all the 0-0 lines, some of which very close in excitation energy
 to the level emitting the 2.12\,$\mu$m line. Hence, we are left with three free parameters: {\it n(H$_2$)}, $\beta$ and the 
fraction of hydrogen nuclei in atomic form, $f$ = n(H)/[2n(H$_2$)+n(H)]. In particular, as shown in Paper\,II in the outflow of L1157, collisions with atomic hydrogen can indeed play an important r\^ole in partially dissociative shocks, given their high efficiency in exciting H$_2$. 
To evaluate the effect of atomic collisions, we have considered two cases, fully molecular gas ($f$=0) and part of
hydrogen in atomic form ($f$=0.5/$f$=0.9); the results for $f$=0 are shown in Figures \ref{L1448_NLTE}-\ref{N2071_NLTE}.\\
In L1448 (Figure \ref{L1448_NLTE}), we generally find higher densities in the nothern lobe, with a density peak close to the outflow tip and to source IRS3; in this part of the outflow, the gas appears close to thermal equilibrium, with values of {\it n(H$_2$)} close to 10$^7$ cm$^{-3}$. By contrast, in the southern lobe {\it n(H$_2$)} is more constant and never exceeds $\sim$ 1-2 10$^6$ cm$^{-3}$. 

The effect of increasing  the assumed value of $f$ is to decrease the derived H$_2$ density (Paper\,II). We have verified that a decrease by a factor between 2 and 5 is found all over the map if $f$=0.5 and up to a factor of 10 if $f$=0.9. Hence, {\it n(H$_2$)} values shown in Figure \ref{L1448_NLTE} have to be considered as upper limits.

In the right panel of Figure \ref{L1448_NLTE} we show the map of the index $\beta$. We fit values typically between 3.5 and 4.5. These values 
are those expected in bow-shocks where a small fraction of H$_2$ is dissociated: indeed a value of $\sim$ 3.8 is predicted for
a single parabolic bow-shock where the apex velocity is above the limit for H$_2$ dissociation; yet higher values are expected for a mixture of bow-shocks, including fully non-dissociative shocks. These latter, indeed, are fast enough to excite only the low-lying H$_2$ levels, therefore increasing the slope in the rotational diagram of the function $dN$ vs, $T$ (Neufeld \& Yuan 2008). We find values of $\beta$ that progressively become higher with increasing distance from the central source: this could result from an increasing number of unresolved bow-shocks with different apex velocities. This result is supported by echelle spectroscopic observations (Davis \& Smith 1996) which evidence very wide (80-100 km s$^{-1}$), typically bow-shaped profiles.    

The density map of BHR71 is presented in Figure \ref{BHR71_NLTE}. A spread of about an order of magnitude, from  
few $\times$ 10$^6$ cm$^{-3}$ to more than 10$^7$ cm$^{-3}$ is found; in general, larger densities are found in the southern lobe, even if some enhancements up to 10$^7$ cm$^{-3}$ are found at the SiO knot and in HH\,320. As in L1448, we obtain a decrease in {\it n(H$_2$)} of up a factor of 5/10 if $f$=0.5/0.9 is taken.
The index $\beta$, presented in the right panel, is generally lower in the northern lobe, where values exceeding 4.0 are more rarely found; in the southern lobe, $\beta$ ranges between 3.5 and 4.5, with the highest values found at the apex shocked region.\\

NGC2071 (Figure \ref{N2071_NLTE}) is the object where we find the highest densities, close to LTE along both the
outflow lobes; indeed densities higher than 2 10$^6$ cm$^{-3}$ are derived also if $f$=0.9 is assumed. The estimated value of
$\beta$ is always larger than 4, with a patchy behaviour that argues for the presence of multiple unresolved bow shocks along the line of sight and/or in the spatial resolution element.

\subsubsection{Analysis of H$_2$ ro-vibrational lines in selected regions\label{sec:sec3.2.2}} 
In the previous setions we analysed 2-D maps of the shock parameters, basing on the {\it Spitzer} and 1-0\,S(1)
line images. Now we will focus on few selected regions, i.e. the 'OF2' position in L1448 (Dionatos et al. 2009) and HH\,320 and HH\,321 in BHR71, labelled in Figure\,\ref{L1448_images} and Figure\,\ref{BHR71_images}.
In these regions a number of H$_2$ ro-vibrational lines have been detected by means of low-resolution spectroscopy between 1.0-2.5\,$\mu$m (Dionatos et al. 2009, Giannini et al.2004). We remark that, in addition to further constrain the physical conditions of these three particular regions, the inclusion  in the rotational diagram of other NIR lines gives us the opportunity to check how much the determination of the hydrogen density depends on the number of the involved lines and hence to validate (or not) the approach described in the previous section, where maps of {\it n(H$_2$)} were constructed relying on the 1-0\,S(1) line only. 

The inter-calibration factor between {\it Spitzer} images and long-slit data has been found by scaling the line flux of the 2.12\,$\mu$m line measured in the slit to the photometry on the 2.12\,$\mu$m image computed on the same areas as in the {\it Spitzer} images (see Table \ref{tab:tab1}). The same factor has been applied to the fluxes of other NIR lines, by assuming the line ratios with the 2.12\,$\mu$m measured with the spectroscopy. For HH\,320 and HH\,321, we have taken an average of the fluxes measured over the sub-structures (HH\,320A/B and HH\,321A/B) that appear resolved at the spatial resolution of the long-slit observations. We have considered lines with S/N larger than 5; this implies that lines with upper vibrational quantum number (v$_{up}$) up to 2 and 3 are considered for L1448 (OF2) and BHR71 (HH\,320/321), respectively.\\
With respect to the NLTE approach described in the previous section, a further modification has to be introduced when lines with higher vibrational number are included in the model. Indeed these lines probe gas at temperatures significantly higher than the pure rotational lines, at which the OPR is expected to have reached the equilibrium value of 3. Indeed, no deviations from this
value are recognizable in the rotational diagrams of the NIR lines in HH\,320 and HH\,321 (Giannini et al. 2004). Consequently, a para-to-ortho conversion time as a function of temperature has to be included in the fitting procedure, so as to allow lines excited at different temperatures to have different OPRs. Following the approach described in Paper\,I and Paper\,II, we have considered the variation of the OPR as a function of the temperature from a starting value OPR$_0$ for a gas that was originally heated at a temperature
$T$ for a time $\tau$. Assuming that the efficiency of the para-to-ortho conversion is regulated by the efficiency of the collisions with atomic hydrogen, the dependence of OPR with temperature can be expressed as: 

\begin{equation}
\frac{OPR(\tau)}{1+OPR(\tau)}  = \frac{OPR_0}{1+OPR_0}\,e^{-n(H)k\tau} + {\frac{OPR_{LTE}}{1+OPR_{LTE}}}\,\left(  1 - e^{-n(H)k\tau}\right) 
\end{equation}

\begin{figure}
\includegraphics[angle=0,scale=0.85]{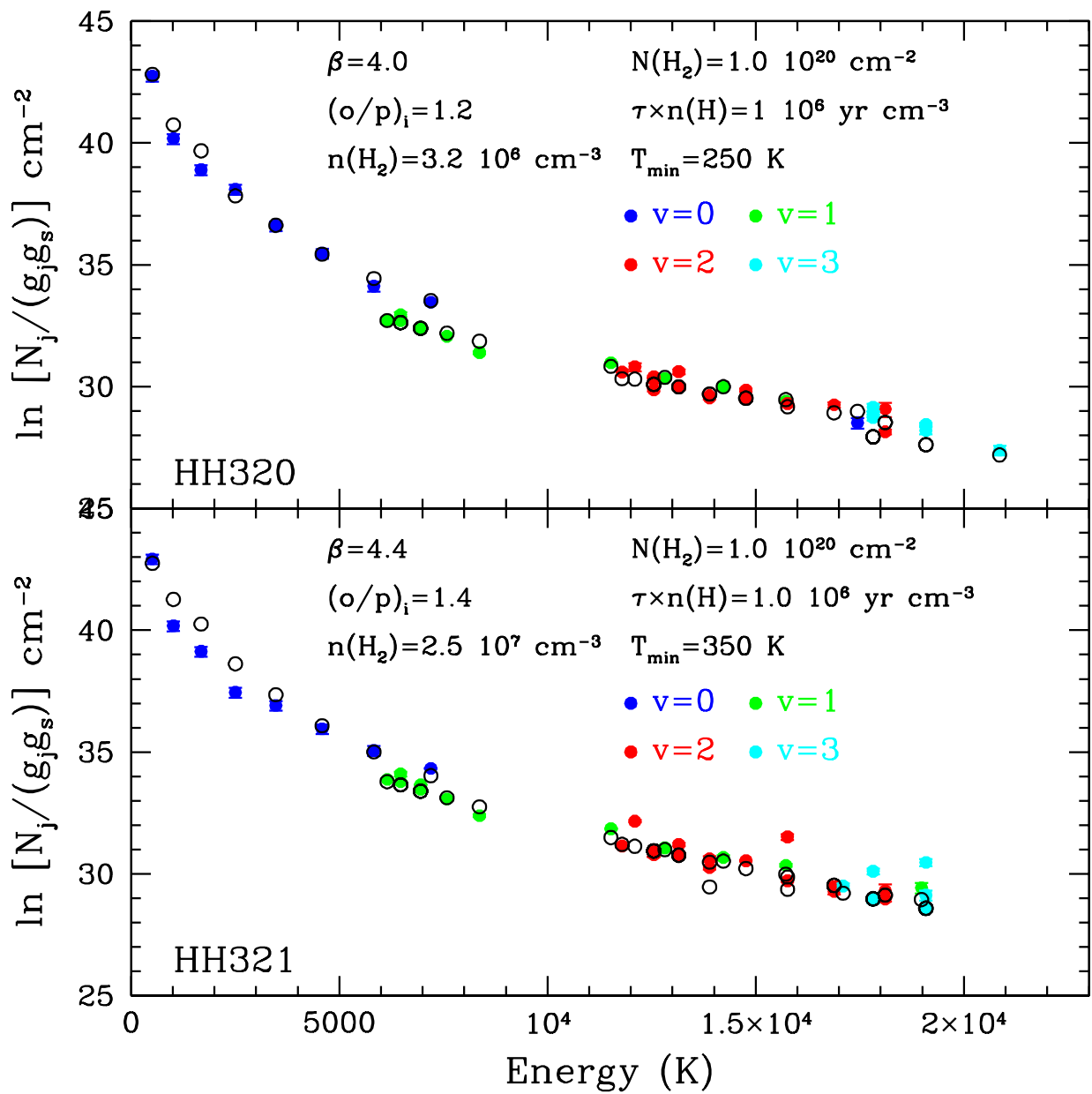}
\caption{Rotational diagram of HH\,320 (top panel) and HH\,321 (bottom panel). Filled and open circles indicate observational and fitted data, respectively; lines coming from different vibrational states are indicated with different colours. The parameters of the best fit through the data is indicated as well.}
\label{HH320}
\end{figure}

In this expression, $n(H)$ is the number density of atomic hydrogen and OPR$_{LTE}$ is the ortho-to-para ratio equilibrium value.
The parameter $ k $ is given by the sum of the rates coefficients for para-to-ortho conversion ($ k_{po} $), estimated
as 8$ \times $10$^{-11}$exp(-3900/T) cm$ ^{3} $\,s$ ^{-1} $, and for ortho-to-para conversion $ k_{op} \sim k_{po}/3 $
(Schofield 1967). Thus the dependence of the OPR on the temperature is implicitly given by the dependence on T of the
$ k $ coefficient. In practice, the inclusion of the NIR lines in our fit introduces as further free parameters both the coefficient $\it{K}$= $\it{n(H)}\tau$ and the initial value OPR$_0$.\\
In Figure \ref{HH320}, we show the best fit we obtain for the HH objects in BHR71. First, we note that {\it n(H$_2$)} differs by about an order of magnitude between HH\,320 and HH\,321, with values  within a factor two with respect to those reported in Figure \ref{BHR71_NLTE}. 
From the ratio  {\it N(H$_2$)}/{\it n(H$_2$)} we can estimate the H$_2$ cooling length: this is $\sim$ 2 10$^{13}$ cm and 5 10$^{12}$ cm
for HH\,320 and 321, respectively.  Both these determinations are well below the cooling lengths predicted by non-dissociative, C-type shock models, which are never less than 10$^{15}$ cm (Neufeld et al. 2006), while are in better agreement with predictions for slow, dissociative shocks ($v_{shock}$ $\approx$ 20-25 km s$^{-1}$, Wilgenbus et al. 2000). 
As far as $\beta$ is concerned, and again with reference to Figure \ref{BHR71_NLTE}, we find a good agreement in HH\,320 and a value a bit higher in HH\,321: we ascribe this discrepancy to the poorer quality of the fit (with respect to HH\,320). 
Other two parameters ({\it N(H$_2$)} and {\it T$_{min}$}), are fully consistent with those originally found by assuming LTE. In regard to the newly determined parameters, we consistently find initial values of the OPR of 1.2 and 1.4 in HH\,320 and in HH\,321, respectively, i.e.  significantly lower than the values between 2 and 2.5 of the present OPRs in the same positions (see Figure \ref{BHR71_NLTE}). The parameter $\it{K}$= $\it{n(H)}\tau$ is fitted to the value 1 10$^6$ yr cm$^{-3}$ for both the
HH objects. Interestingly, we can remove the degeneracy between $\it{n(H)}$ and $\tau$, by assuming for this latter the age of the shock\footnote{The shock age is defined as the ratio between the length of the shock wave (roughly comparable with the dimension the shocked region), and the shock velocity.}
determined by Gusdorf et al.(2011); the best fit gives an age of 815 yr and 700 yr for HH\,320 and HH\,321, respectively. These imply, for HH\,320, $\it{n(H)}$ = 1.2 10$^3$ cm$^{-3}$, and $\it{f}$ $\sim$ 4 10$^{-4}$, and, for HH\,321 $\it{n(H)}$ = 1.4 10$^3$ cm$^{-3}$, and $\it{f}$ $\sim$ 5 10$^{-5}$. Hence, we conclude that most of H$_2$ remains in molecular form. This result agrees both with the non-detection of H$\alpha$ (Corporon \& Reipurt 1997) and $[FeII]$ emission (Giannini et al. 2004), and also with the results coming from shock models, which are consistent with low-velocity, partially dissociative shocks (Gusdorf et al., 2011). Interestingly, significantly higher values of $\it{f}$ ($\sim$ 0.3) are found in Paper\,II in L1157, where evidence for a large fraction of ionized gas has been recognized (Caratti o Garatti et al. 2006)\\

\begin{figure}
\includegraphics[angle=0,scale=0.85]{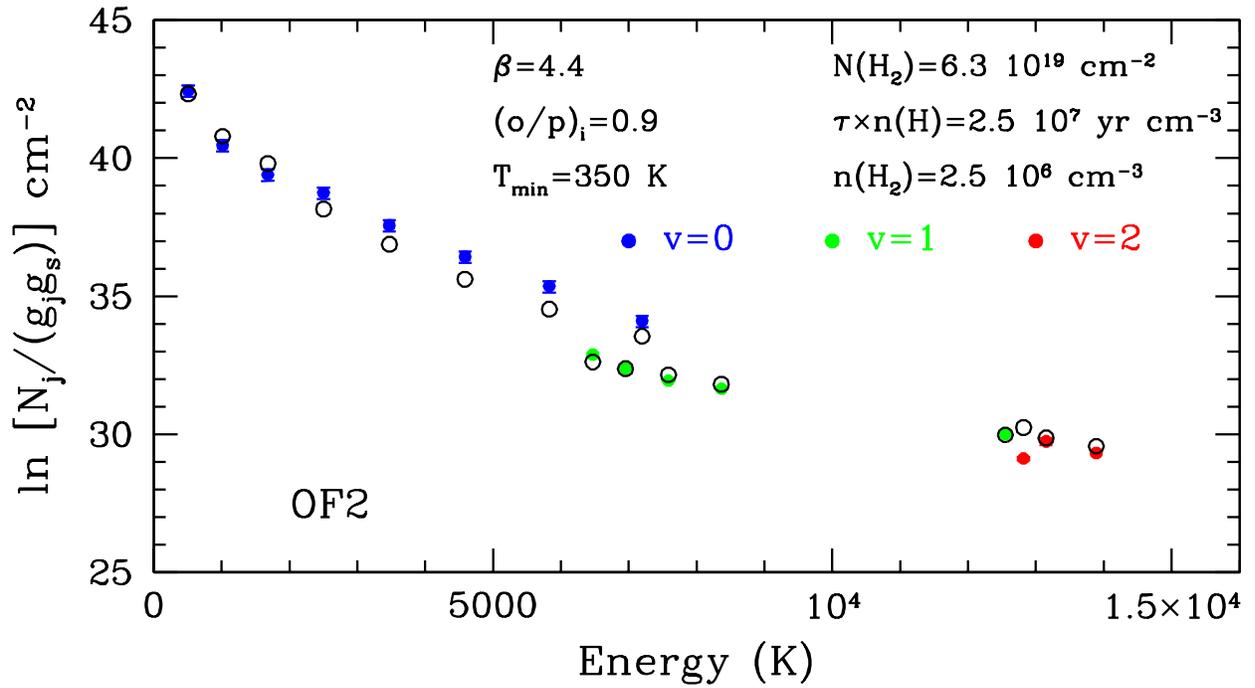}
\caption{Rotational diagram of OF2 position in L1448. Symbols have the same meaning as in Figure \ref{HH320}.}
\label{OF2}
\end{figure}

In Figure \ref{OF2}, we present the rotational diagram for the OF2 position in L1448. Although fewer rovibrational lines are observed here, we successfully constrain all the parameters in the fit, finding a good agreement with the values from the LTE model. In this case, we find a very low value of OPR$_0$ (0.9). For comparison, initial conditions very close to those of the cold molecular cloud, where H$_2$ is expected to be almost all in the para form, have been invoked by Maret et al. (2009) to explain the extremely low inital OPR ($\sim$ 0.5) measured toward NGC1333. Conversely, the higher values found in BHR71 could indicate that intermittent shock events have occurred, with a temporal interval shorter than the time needed to H$_2$ to cool enough to convert all the ortho states in the para form.
\\

\section{Analysis of the HD emission\label{sec:sec4}} 
In L1448 and in NGC2071, we detected two HD lines, the 0-0 R(3) transition at 28.50 $\mu$m  and the 0-0 R(4) transition at 23.03 $\mu$m, in two distinct positions along the outflows (see Figures \ref{L1448_images} and \ref{N2071_images}). The observation of HD lines in NGC2071 was presented
previously by Melnick et al. (2008), who used them to infer a fractional abundance of HD of $\sim$ 1-2
10$^{-5}$. Here, we will use the HD detections to derive the physical conditions of the emitting gas independently from the H$_2$ emission. 

\begin{figure}
\includegraphics[angle=0,scale=0.85]{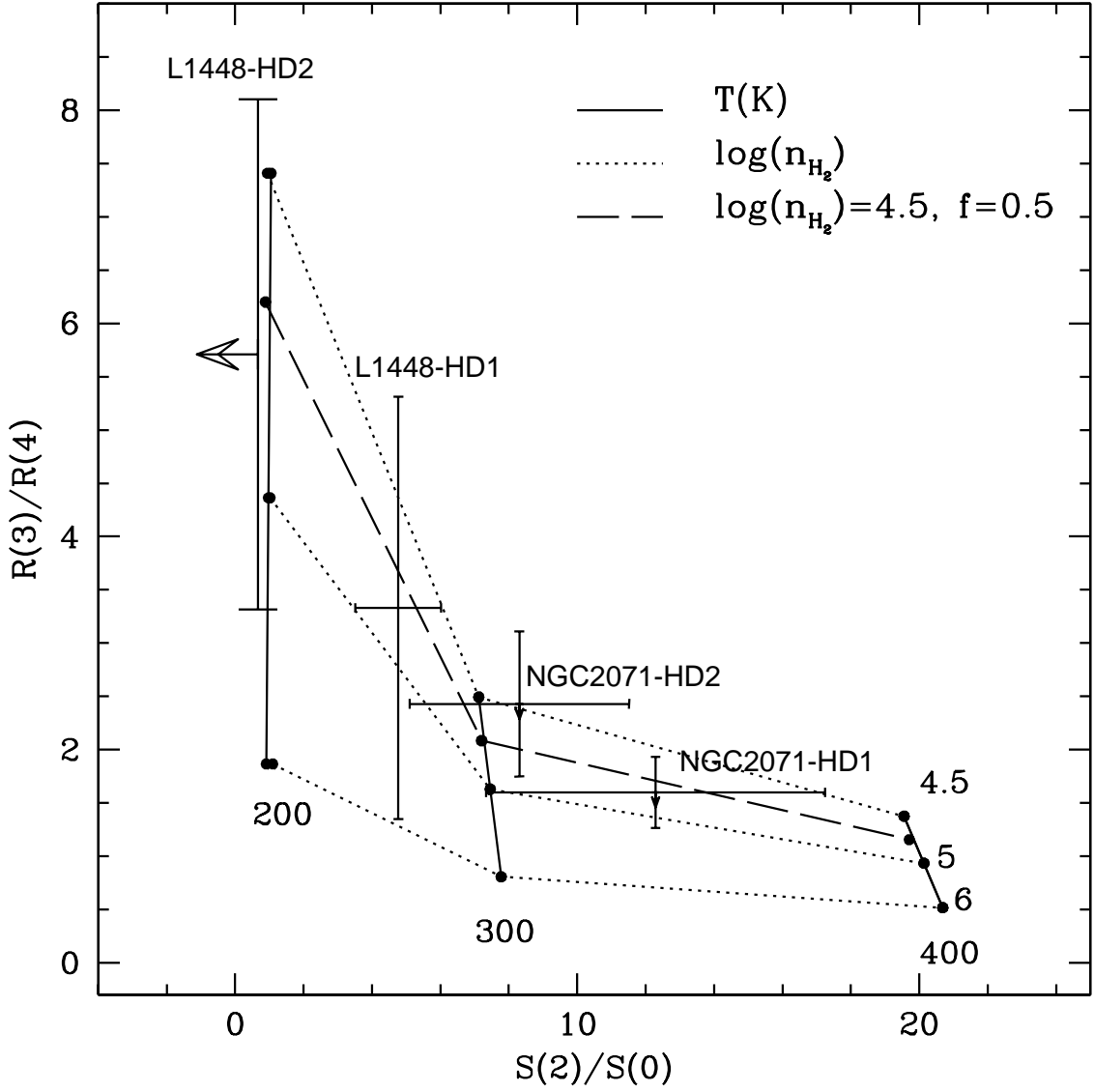}
\caption{Diagnostic diagram of the ratio HD R(3)/R(4) against H$_2$ S(2)/S(0), as a function of a grid of temperature (200 $< T <$ 400 K, continuos lines) and density ($10^{4.5} < n(H_2) < 10^6$ cm$^{-3}$, dotted lines). The effect of introducing atomic hydrogen as a collision partner is shown with a long-dashed curve, for log({\it n(H$_2$)})=4.5 and $f$=0.5: this curve coincides with the curve with log[{\it n(H$_2$)}]=4.7 and $f$=0. Hence, an increase of $f$ implies lower {\it n(H$_2$)}.}
\label{HD}
\end{figure}

The R(3) and R(4) lines originate from levels with excitation energy of 532 and 883 cm$^{-1}$, lower than the level v=0, J=4 (1168 cm$^{-1}$) emitting the S(2) line. Hence it is reasonable to assume that the bulk of the observed HD emission originates from the same (cold) gas component that emits also the H$_2$ S(0)-S(2) lines. The closeness of the excitation energy of the two HD lines makes them not particurlarly suitable for determining the temperature; however, unlike the H$_2$ 0-0 lines, they are characterized by higher critical densities (around 5 10$^4$ - 10$^5$ cm$^{-3}$ at 200 K) and hence they can efficiently probe the molecular gas density. To this end, we have applied a 8 level NLTE model for HD emission, including collisions with H$_2$ and atomic hydrogen. Collisional and radiative coefficients were taken from Flower et al. (2000). As output, we constructed the diagnostic diagram shown in Figure \ref{HD}. We plot  the ratio of the HD lines R(3)/R(4) against that of the H$_2$ lines S(2)/S(0) over a grid of temperature (from 200 K to 400 K, continuous lines) and molecular hydrogen density (from 3 10$^4$ to 10$^6$ cm$^{-3}$, dotted lines). With a long dashed line we show how the  R(3)/R(4) ratio changes in the case {\it n(H$_2$)}=3 10$^4$ cm$^{-3}$ and  $\it{f}$ = 0.5. Superimposed on the diagram are the ratios measured in the HD spots of L1448 and NGC2071 (labelled in Figure\,\ref{L1448_images} and Figure\,\ref{N2071_images}).

\begin{figure*}
\includegraphics[angle=-90,scale=0.7]{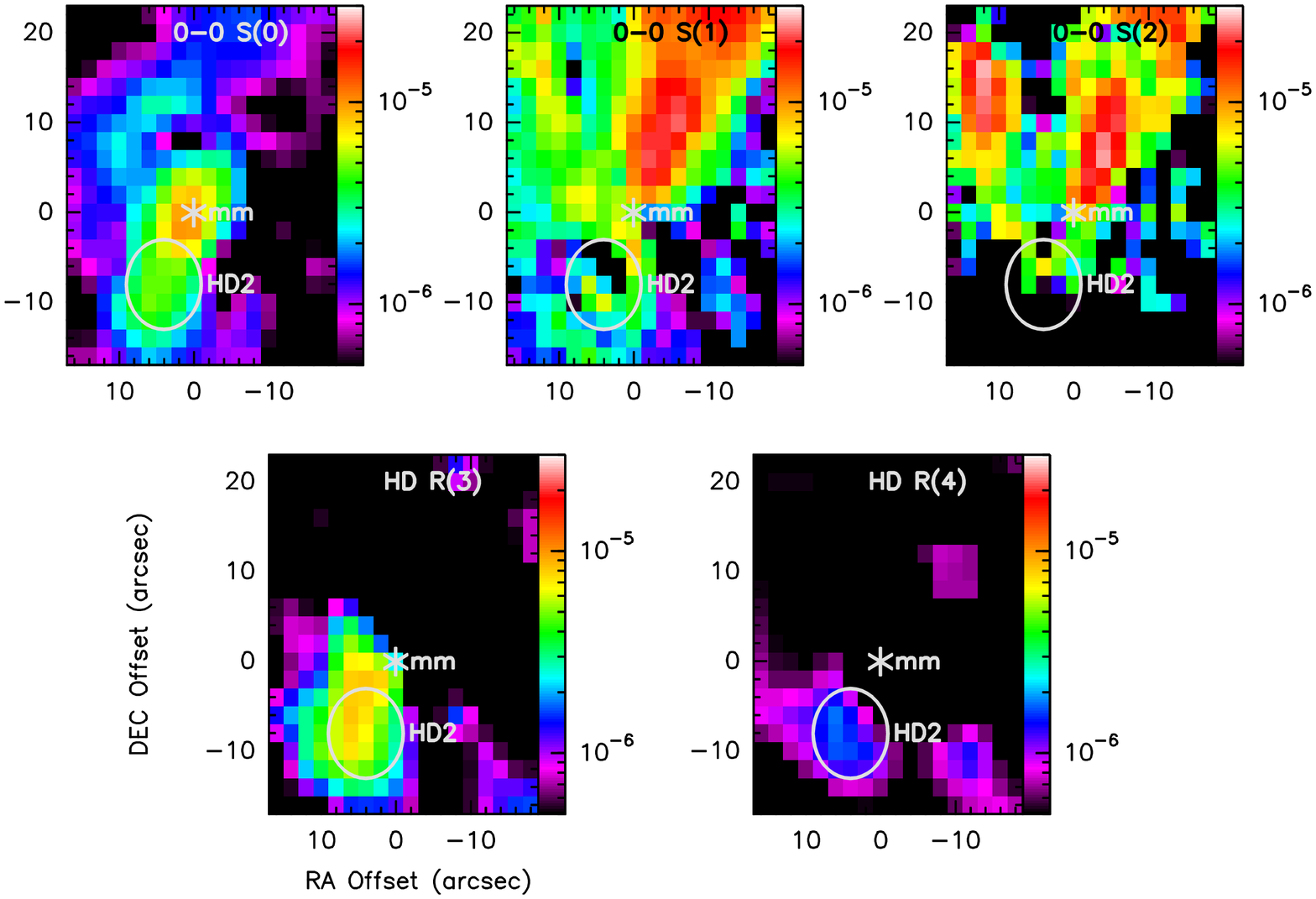}
\caption{L1448 images of H$_2$ S(0)/S(1)/S(2) and HD R(3)/R(4) in a 10 $\times$ 20 arcsec box around L1448-mm. The same emission scale is used for 
an easy by-eye comparison of the relative brightness.} 
\label{L1448_HD}
\end{figure*}

From this diagram we note that: 
\begin {itemize} 
\item[-] the density is lower than few 10$^5$ cm$^{-3}$ in NGC2071 and 10$^6$ cm$^{-3}$ in L1448. This is in evident contrast with the maps we have derived with the 2.12\,$\mu$m line, where {\it n(H$_2$)} ranges between 10$^6$ and more than 10$^7$ cm$^{-3}$. The two results cannot be reconciled even if a very large fraction of atomic hydrogen is assumed, since the effect would be to further decrease {\it n(H$_2$)} (see also \textsection\ref{sec:sec3.2.1} and the caption of Figure \ref{HD}). More likely, the cold gas component, namely that traced by the low-$J$ H$_2$ lines, is at density lower than the warm gas traced by the rovibrational lines. This implies that, on the one hand, a density stratification of the molecular gas does occur, and, on the other hand, that the densities derived with the 2.12 $\mu$m have to be considered as an average over all the temperature components, and as stringent upper limits for the cold gas component.
\item[-] interestingly, the data obtained at HD2 in L1448 indicate a gas temperature lower than 200 K. This result comes directly from the absence
of lines with $J\ge$ 1 in this position, as evidenced in Figure\,\ref{L1448_HD}. We estimated the H$_2$ column density at HD2 by considering the detection of the S(0) and the 3\,$\sigma$ upper limit on the S(1) line. We find {\it N(H$_2$)} $\ge$ 5 10$^{20}$ cm$^{-2}$, larger than other estimates along the L1448 flow (Figure\, \ref{L1448_all}): hence we can explain the detection of HD lines at the HD2 location with a local enhancement of the column density of the coldest gas component close to L1448-mm.
\item[-] in the other three spots of HD (HD1 in L1448 and HD1/2 in NGC2071), the gas temperature exceeds 250 K: according to our NLTE model these conditions are favourable for having bright HD lines. Possible local enhancements of the HD abundance can also explain the detection of the HD spots: this topic will be treated in dedicated paper (Yuan et al. in preparation).   
\end {itemize}

\section{Comparison with results of Paper\,I and II\label{sec:sec5}}
The first aim of this section is the comparison of our fit results with those obtained on the {\it same} sources in Paper\,I, which were however obtained {\it by averaging the observed line intensities all over the flow}. In that work, the H$_2$ rotational lines have been fitted in a NLTE framework, taking as free parameters $M(H_2)$, $\beta$, $n(H_2)$, OPR$_0$, and $n(H)\tau$.
With reference to Table\,\ref{tab:tab2} of the present work and Table\,3 of Paper\,I, we first note that marginal differences exist among {\it N(H$_2$)} and {\it $M(H_2)$} derived in the LTE and NLTE treatments. In general we find sligthly higher values (within a factor of two), likely arising from having included in our fits the S(0) line, which probes the emission of the diffuse molecular gas.
In the case of the parameters we determine in NLTE, i.e. $\beta$ and {\it n(H$_2$)}, the two approches give substantally different results. 
The best fits of Paper\,I give $\beta$ between 2.3 and 3.3 and log({\it n(H$_2$)})$\sim$ 3.8. By contrast, we get $\beta$ always larger than 3.8, and log({\it n(H$_2$)})$\ga$ 6; moreover, though at a lesser extent, {\it n(H$_2$)} remains larger than the Paper\,I determinations even in locations where we are able to remove the degeneracy with $\beta$. We ascribe this major discrepancy to the sensitivity of the 0-0 lines only to the low-density components, a problem that has been overcome in the present work thanks to the inclusion of the 2.12\,$\mu$m line in the analysis. Noticeably, for the same reason, in the northern lobe of NGC2071 we probe densities at least an order of magnitude higher than those estimated by Melnick et al. (2008) with the {\it Spitzer} lines. 

Finally, it is interesting to compare the parameters of the three outflows considered here with those obtained in Paper\,II on L1157. These are summarized in the last row of Table \ref{tab:tab2}. The values fitted in L1157 are of the same order of magnitude as in the other objects, arguing for very similar physical conditions of the molecular gas in all these outflows from Class 0 sources. We only note a lower value of the initial ortho-to-para ratio, that is on average $\sim$ 1.8, and a significantly higher $\beta$ (up to 5.5). While the former might indicate a lower temperature of the gas prior to the passage of the shock, the latter might not reflect intrinsic shock conditions, but rather be due to the LTE assumption under which $\beta$ was computed in Paper\,II.

\section{Energetics\label{sec:sec6}}
\input{tab3.tex}

According to models for non-dissociative shocks (e.g. Kaufman \& Neufeld 1996), H$_2$ emission represents one of the major channels for radiating
away the mechanical energy of the flow. For example, for pre-shock densities lower than 10$^5$ cm$^{-3}$ and shock velocities larger than 20 km s$^{-1}$, the fraction of radiation in form of H$_2$ lines is between 40\% and 70\%. 
The luminosity of H$_2$ S(0)-S(7) rotational lines for the three outflows presented here has been estimated in Paper\,I (their Table\,3). Other contributions from bright H$_2$ lines can be evalutated by considering detailed predictions of shock models: for example Smith (1995), predicts at least ten near-infrared lines (mainly coming from the v=1 level) with intensity comparable to that of the 2.12\,$\mu$m; since from our images we measure L(2.12) $\sim$ 0.1-0.2 L(0-0), we can roughly estimate the total H$_2$ luminosity by simply doubling that of the 0-0 lines. For example, we get in this way L(H$_2$) determinations for L1448 and BHR71 in agreement within a factor of two with those estimated by Caratti o Garatti (2006) in LTE approximation.
   
Other important contributions to the total cooling come from species emitting in the far-infrared (Kaufman \& Neufeld 1996). These have been measured in shocks from Class 0 sources by Giannini, Nisini, \& Lorenzetti (2001); basing on ISO observations they found that the FIR cooling (L(FIR)=L(OI)+L(CO)+L(H$_2$O)+L(OH)) is between 10$^{-3}$ and 10$^{-1}$ L$_{\odot}$, the dominant cooling species being CO and H$_2$O. 

In L1448, the L(0-0) luminosity of the northern lobe is 0.029 L$_{\odot}$, i.e. more than twice in the southern lobe (recall, however, that the most distant bow-shock of the southern lobe was not mapped). This difference 
can be partially due to the larger extension of the northern lobe (by about a factor 1.5), but likely also reflects the presence of internal 
shocks caused by the interaction of the two outflows coming from L1448-mm and the triple system L1448-IRS3. This is also demonstrated by the higher densities measured, on average, in the northern lobe, and, in particular, at the peak N: here we register the highest luminosities, also in comparison with peak S in the southern lobe (see Table \ref{tab:tab1}).     
The higher luminosity of the northern lobe is evident also in the FIR cooling, where L(FIR) is 0.076 L$_{\odot}$ (against 0.047 L$_{\odot}$ in the southern lobe), even if the largest value is measured close to the central source where L(FIR)=0.23 L$_{\odot}$ (Nisini et al. 2000). \\
As shown in Table\,\ref{tab:tab3}, summing up all the contributions (H$_2$ + FIR), we get an estimate of the total shock cooling, L$_{cool}$ $\sim$ 0.44 L$_{\odot}$. Since we are not able to distinguish the single contributions of each of the two outflows, we compare this quantity with the sum of the bolometric luminosities of the two driving sources: these are 11 and 9  L$_{\odot}$ for L1448-IRS3 and L1448-mm, respectively (Andr\'e et al. 2000). The so derived L$_{cool}$/L$_{bol}$ ratio is in agreement with other determinations for Class 0 sources (Giannini, Nisini, \& Lorenzetti, 2001). L$_{cool}$ is also comparable with the most recent estimate of the kinetic energy of the molecular outflow (Curtis et al. 2010). Hence, the energy released in the multiple shocks inside the jet is shared out roughly in equal parts between line radiation and mechanical energy of the outflow.

In the case of BHR71, we have computed separately the L(0-0) luminosity for the northern lobes of the two outflows driven by the two sources IRS1 and IRS2. These are L(0-0)= 0.019 L$_{\odot}$ and 0.008 L$_{\odot}$, respectively. The two southern lobes cannot be separated observationally;
if we assume the same proportions as for the northern lobes we get for the IRS1 (IRS2) outflow L(0-0)= 0.015 (0.006) L$_{\odot}$. 
With the same prescriptions adopted for L1448, the rough estimate of the total H$_2$ cooling for the two outflows is 0.068 L$_{\odot}$ and 0.028 L$_{\odot}$. Since there are no estimates for the FIR cooling, we assume L$_{cool}$ ranging between L(H$_2$) and 4$\times$ L(H$_2$),
under the assumption that H$_2$ makes at least a 25\% contribution to L$_{cool}$.
With the new determinations of the bolometric luminosities of the two sources, based on interferometric observations, of  13.5 and 0.5 L$_{\odot}$ (Chen et al. 2008), we derive L$_{cool}$/L$_{bol}$=5 10$^{-3}$-2 10$^{-2}$ and 5.6 10$^{-2}$-2.2 10$^{-1}$ for IRS1 and IRS2, respectively. 
(see Table\,\ref{tab:tab3}).

Our estimate of the H$_2$ cooling in NGC2071 is certainly underestimated, since our observations cover the south lobe only partially. The separate computation of L(0-0) in the two lobes gives L(0-0)=0.57 L$_{\odot}$ and 0.19 L$_{\odot}$ for the northern and southern lobe, respectively. If we attribute the same luminosity for each lobe, we get L(0-0)$\sim$1.14 L$_{\odot}$, and L(H$_2$)$\sim$2.3 L$_{\odot}$. Again, if we assume that H$_2$ contributes at least 25\% to the total cooling, a reasonable range for L$_{cool}$ is between 2.3 and 10 L$_{\odot}$. We compare this rough
evaluation with the 520 L$_{\odot}$ total luminosity of the cluster, composed of eight distinct infrared sources (Butner et al. 1990, Walther et al. 1993). The ratio L$_{cool}$/L$_{bol}$ is within the typical range observed for outflows coming from Class 0 sources (see Table\,\ref{tab:tab3}). Finally, our estimate of L$_{cool}$ falls between the two more recent estimates of the kinetic luminosity of the high velocity gas of 1.6 L$_{\odot}$ (Stojimirovi\'c, Snell, \& Narayanan 2008) and 180 L$_{\odot}$ (Buckle et al. 2010). These estimates are so discrepant that it is impossible to perform a meaningful comparison with our L$_{cool}$ determination.

\section{Conclusions\label{sec:sec7}}
We have presented the H$_2$ pure rotational lines from S(0) to S(7) observed with IRS-{\it Spitzer} in three outflows from Class 0
protostars: L1448, BHR71, and NGC2071. We have analyzed the data by assuming a gas temperature stratification that follows a power law distribution. The mid-IR data have also been combined with near-infrared H$_2$ rovibrational lines and mid-IR HD lines detected in few spots to derive the physical parameters under a NLTE approach. Our main results are summarized in the following:
\begin{itemize}

\item[-]We have obtained detailed maps of gas column density, OPR and temperature assuming the LTE approximation.    
All the flows present similar characteristics in regard to the covered ranges of these parameters, and the derived values are similar to the same
quantities measured in Paper\,II in L1157. Typically the column density ranges between  1-7 10$^{20}$ cm$^{-2}$, corresponding to a mass of gas of the order of 10$^{-2}$ M$_{\odot}$. The only exception is NGC2071, which is definitely more massive than the other two (M$\sim$ 0.3 M$_{\odot}$).
 
\item[-] A large range of temperatures, between few hundreds to few thousands of Kelvin, contribute to the H$_2$ emission, in all three of the flows. The hottest components are found typically along the jet axis and in warm spots, corresponding  to Herbig-Haro objects or to small regions where HD emission is also detected.

\item[-] Significant deviations from the equilibrium value of the ortho-to-para ratio are evident in all the three outflows. On average we find values around 2.2, with a large range that goes from less than one to three. However, some trends can be identified, with the northern lobes of  L1448 and BHR71 closer to the equilibrium than in the south, and with a decrease in NGC2071 going from the internal toward the apexes of the flow. Generally speaking, higher OPRs are found at the location of temperature peaks, even if low OPR values can sometimes be found also in regions with relatively high temperatures, as for example in the southern lobe of BHR71. These apparently contrasting results might be explained if the time elapsed from the last shock event were no long enough to allow reaching the low OPR values.

\item[-] H$_2$ rovibrational lines, and in particular the 1-0\,S(1) line at 2.12\,$\mu$m, have been used in conjunction to the rotational lines to relax the LTE approximation, and to derive detailed maps of the post-shock molecular gas density and the power law index that regulates the temperature stratification. As far as this latter parameter is concerned, we find values always larger than 3.8, in agreement with shock models for multiple C-types bow shocks with apex velocities able to partially dissociate H$_2$. 

\item[-] H$_2$ densities are roughly between 10$^6$-10$^7$ cm$^{-3}$; these values, however, can decrease by up to a factor of ten if a significant fraction of atomic hydrogen exists; moreover, from the analysis of the HD emission we demonstrate that indeed a density stratification does exists,
with the lowest densities corresponding to the lowest temperature components.

\item[-] In a few selected regions we were able to construct rotational diagrams up to excitation energies of $\sim$ 20\,000 K and $v \le$ 3. From these diagrams we are able to estimate the time needed for the para-to-ortho conversion as a function of the fraction of the atomic hydrogen. In HH\,320 and HH\,321, in particular, we used the results of recent shock models to estimate that fraction, finding that hydrogen is almost all in molecular form. This result, although regarding small regions inside the flow, gives a different picture of that found in L1157, where the fraction of atomic hydrogen has been found to be $\sim$ 0.1-0.3.

\item[-] We have studied in detail a region, about 20$^{\prime\prime}$ in length, close to the L1448-mm source. Here very different conditions seem
to occur in the two lobes: in the northern one all the H$_2$ rotational lines are detected, indicating a temperature $\sim$ 300 K; conversely in the south lobe only the fundamental line is detected, along with the HD lines R(3) and R(4). These findings are consistent with a gas temperature
lower than 200 K and a local enhancement of the column density: we expect consequently that a different chemistry could have taken place in the two lobes.

\item[-] An estimate of the radiative cooling of H$_2$ was derived and compared, when available, with the cooling of abundant molecules emitting in the far-infrared. The estimated total cooling, if compared with the kinetic energy of the swept-out outflow, supports the scenario in which the shock from which H$_2$ is emitted is also capable of accelerating the molecular outflow.   

\end{itemize}

Finally, we note that the objects analyzed in the present paper are going to be observed with the far-infrared and submillimetre telescope {\it Herschel}: the map parameters we have provided here will be essential for the interpretation of the forthcoming data, and will be helpful for deriving a more unified view of the physical and chemical processes governing the ongoing shocks.  

\section{Acknowledgements}
We are grateful to Carolyn McCoey for her useful comments to the paper;  
we also thank Jochen Eisl\"offel for having provided us with the 2.12\,$\mu$m image of NGC2071. 
This paper is based on observations made with the Spitzer Space Telescope, which is operated by the Jet
Propulsion Laboratory, California Institute of Techonology under a contract with NASA. TG, BN and SA acknowledge 
financial contribution from the agreement ASI-INAF 1/009/10/0.

\end{document}

%% file: tab1.tex
\begin{deluxetable}{cccccccc}
\tabletypesize{\scriptsize} \tablewidth{0pt}
\tablecaption{H$_2$ photometry in selected areas.\label{tab:tab1}} 
\tablehead{
Transition& $\lambda (\mu m)$ &  Module   & \multicolumn{5}{c}{Brightness$^a$ (10$^{-5}$ erg cm$^{-2}$ s$^{-1}$ sr$^{-1}$)}}
\startdata
          &                   &           & \multicolumn{5}{c}{L1448$^b$}\\
\cline{1-8}
          &                   &           &    N      &   OF2     &   HD1    &   HD2  &  S	 \\
          &                   &           & (-32,+66) & (-12,+34) & (-30,+64)& (+4,-8)& (+26,-75)\\
\cline{4-8}
0-0S(0)	  &    28.21          &    LH     &   0.5     &  0.2      &   0.9    &   0.4  &  0.3     \\
0-0S(1)   &    17.03          &    SH     &   4.4     &  2.3      &   4.4    &$<$0.2  &  2.6     \\
0-0S(2)   &    12.28          &    SL     &   4.3     &  2.4      &   4.3    &$<$0.3  &  2.8     \\
0-0S(3)   &     9.66          &    SL     &  16.7     & 12.0      &  19.8    &$<$2    & 10.5     \\
0-0S(4)   &     8.02          &    SL     &   7.4     &  7.9      &   8,3    &$<$3    &  5.9     \\
0-0S(5)   &     6.91          &    SL     &  20.1     & 25.1      &  16.2    &$<$2    & 18.1     \\
0-0S(6)   &     6.11          &    SL     &   5.3     &  7.7      &   4.5    &$<$2    &  5.4     \\
0-0S(7)   &     5.51          &    SL     &  10.7     & 11.9      &   5,5    &$<$2    &  9.5     \\
1-0S(1)   &     2.12          &    -      &   1.4     &  1.5      &   1.2    &$<$0.1  &  0.6     \\
HD R(3)   &    28.50          &    LH     &    -      &   -       &   1.0    &  0.6   &  0.3     \\
HD R(4)   &    23.03          &    LH     &    -      &   -       &   0.3    &  0.1   &  0.1     \\
\cline{1-8} 
          &                   &           &\multicolumn{3}{c|}{BHR71$^c$}& \multicolumn{2}{c}{NGC2071$^d$}\\
\cline{1-8}
          &                   &           &  HH320    &  HH321    & \multicolumn{1}{c|}{SiO knot} & HD1        &   HD2     \\
          &                   &           & (-31,+38) & (+3,-45)  & \multicolumn{1}{c|}{(-5,+75)} & (+48,+68)  & (-62,-60) \\
\cline{4-8}        
0-0S(0)	  &    28.21          &    LH     &   0.3     &   0.4     & \multicolumn{1}{c|}{ 0.4}	  &  2.0       &   1.8     \\
0-0S(1)   &    17.03          &    SH     &   2.0     &   2.1     & \multicolumn{1}{c|}{ 2.8}	  & 12.9       &  11.4     \\
0-0S(2)   &    12.28          &    SL     &   2.1     &   2.3     & \multicolumn{1}{c|}{ 3.6}	  & 24.7       &  15.3     \\
0-0S(3)   &     9.66          &    SL     &  15.2     &   6.4     & \multicolumn{1}{c|}{18.8}	  & 55.0       &  34.9     \\
0-0S(4)   &     8.02          &    SL     &   6.0     &   6.6     & \multicolumn{1}{c|}{ 9.7}	  & 46.4       &  31.4     \\
0-0S(5)   &     6.91          &    SL     &  14.7     &  17.0     & \multicolumn{1}{c|}{24.6}	  &124.0       &  85.2     \\
0-0S(6)   &     6.11          &    SL     &   6.9     &   7.9     & \multicolumn{1}{c|}{ 8.8}	  & 27.7       &  19.6     \\
0-0S(7)   &     5.51          &    SL     &  14.0     &  18.5     & \multicolumn{1}{c|}{20.5}	  & 59.2       &  48.9     \\
1-0S(1)   &     2.12          &    -      &   5.3     &  13.5     & \multicolumn{1}{c|}{ 8.2}	  & 28.1       &  21.1     \\
HD R(3(   &    28.50          &    LH     &    -      &   -       & \multicolumn{1}{c|}{   -}     &  0.4       &   0.4     \\
HD R(4)   &    23.03          &    LH     &    -      &   -       & \multicolumn{1}{c|}{   -}	  &  0.2       &   0.1     \\

\cline{1-8}
\enddata
\tablenotetext{a}{Brightness computed over a 5$^{\prime\prime}$ FHWM Gaussian aperture around the indicated position. Absolute
uncertainties are between 20\% and 40\%.}
\tablenotetext{b}{Positions respect to L1448-mm (3h 25m 38.9s 30d 44$^{\prime} 06^{\prime\prime}$).}
\tablenotetext{c}{Positions respect to IRS1 (12h 01m 36.7s -65d 08$^{\prime} 48^{\prime\prime}$.5).}
\tablenotetext{d}{Positions respect to IRS1 (5h 47m 4.8s 0d 21$^{\prime} 42^{\prime\prime}$.8).}
\end{deluxetable}

%% file: tab2.tex
\begin{deluxetable}{cccccccc}
\tabletypesize{\scriptsize} \tablewidth{0pt}
\tablecaption{Parameter statistics$^a$.\label{tab:tab2}} 
\tablehead{
 Object    & N(H$_2$)$^b$          & M(T$>$T$_{min}$)$^b$   & OPR$^b$      & T$_{cold}^b $ & T$_{warm}^b$     & $\beta$$^c$  & n(H$_2)^{c,d}$\\
           & (10$^{20}$ cm$^{-2}$) & (10$^{-2}$ M$_{\odot}$)&              &    (K)        &  (K)             &              & (10$^6$ cm$^{-3}$)}
\startdata
 L1448 	   &   1.1 (0.4-4)         &  1.0                   &2.3 (0.5-2.8) & 310 (270-440) & 1250 (1000-1300) & 4.1(3.5-4.5)  & 0.3-2  \\
 BHR71 	   &   1.0 (0.3-2)         &  0.6                   &2.0 (1.7-3.0) & 310 (250-400) & 1500 (1100-1650) & 3.8(3.0-4.5)  & 2-7.5  \\
 NGC2071   &   7.5 (5-18)          &  26.2                  &2.2 (1.9-2.9) & 240 (180-300) & 1250 (1100-1400) & 4.2(3.6-4.5)  & 6-19   \\
 L1157     &   1.4 (0.5-3)         &   -                    &1.8 (0.4-3.0) & 280 (220-470) & 1230 (1000-1550) & 4.5-5.5$^e$   &  -     \\        
\enddata
\tablenotetext{a}{Parameter value averaged over the entire map along with its range of variation (in parenthesis).}
\tablenotetext{b}{Parameter derived in LTE approximation.}
\tablenotetext{c}{Parameter derived in NLTE approximation.}
\tablenotetext{d}{Values referring to f=0.5 and f=0, respectively.}
\tablenotetext{e}{Obtained under LTE approximation by N10.}
\end{deluxetable}

%% file: tab3.tex
\begin{deluxetable}{cccccc}
\tabletypesize{\scriptsize} \tablewidth{0pt}
\tablecaption{Line Cooling$^1$.\label{tab:tab3}} 
\tablehead{
Object     & L(0-0)         & L(H$_2$)       & L$_{cool}$      & L$_{kin}$     & L$_{cool}$/L$_{bol}$ \\
   -       & (L$_{\odot}$ ) & (L$_{\odot}$)  & (L$_{\odot}$)   & (L$_{\odot}$) & -         }   
\startdata                                         
L1448 	   & 0.044          & 0.09         &  0.44           &  0.33      &  0.022                       \\
BHR71 	   & 0.048          & 0.096        &  0.096-0.36     &      -     &  0.0005-0.02$^a$; 0.06-0.2$^b$\\
NGC2071    & $\sim$1.14     & 2.3          &  2.3-10         &  1.6-180   &  0.0004 -0.02$^{c}$           \\
\enddata
\tablenotetext{1}{References are given in the text}
\tablenotetext{a}{: computed for IRS1; $^b$:computed for IRS2; $^c$:value referring to the whole cluster of 8 sources.}
\end{deluxetable}